\documentclass[a4paper,11pt]{article}
 \usepackage{amsfonts}
 
 \usepackage{graphicx}
 \usepackage{amsmath}
 \usepackage{amssymb}
 \usepackage{latexsym}
 \usepackage[colorlinks]{hyperref}
 \usepackage{color}
 \usepackage{float}
 \usepackage{cite}
 \usepackage{makeidx}
 \usepackage{colortbl}
 \usepackage{csquotes}
 \usepackage[squaren]{SIunits}
 
 \usepackage[textfont={footnotesize,sf},labelfont={color=blue,bf,sf},labelsep=endash]{caption}
 \usepackage[position=top,labelfont={color=blue,bf,sf}]{subfig}
 \usepackage{bm}

 \usepackage[title]{appendix}

\begin{document}
\begin{titlepage}
\setcounter{page}{1}
\renewcommand{\thefootnote}{\fnsymbol{footnote}}

\begin{flushright}
        \end{flushright}

        \vspace{5mm}

\begin{center}
  {\Large{\bf Density of States Analysis of Electrostatic Confinement\\ in Gapped Graphene}} 
\vspace{5mm}

{\bf Ahmed Bouhlal}$^a$, {\bf Abdelhadi Belouad}$^a$,
{\bf Ahmed Jellal}$^{a,b}$\footnote{\sf a.jellal@ucd.ac.ma}
and {\bf Hocine Bahlouli}$^c$

\vspace{5mm}

{$^a$\em Laboratory of Theoretical Physics, Faculty of Sciences, Choua\"ib Doukkali University},\\
{\em PO Box 20, 24000 El Jadida, Morocco}

{$^{b}$\em Canadian Quantum  Research Center,
204-3002 32 Ave Vernon, \\ BC V1T 2L7,  Canada}

{$^c$\em Physics Department,  King Fahd University
	of Petroleum $\&$ Minerals,\\
	Dhahran 31261, Saudi Arabia}

\vspace{3cm}

\begin{abstract}
We investigate the electrostatic confinement of charge carriers in a gapped graphene quantum dot in the presence of a magnetic flux.
The circular quantum dot is defined by an electrostatic gate potential delimited in an infinite graphene sheet which is then connected to 
a two terminal setup. Considering different regions composing our system, we explicitly determine the solutions of the energy spectrum 
in terms of Hankel functions. Using the scattering matrix together with the asymptotic behavior of the Hankel functions for large arguments, 
we calculate 
the density of states  and show that it has an oscillatory behavior with the appearance of resonant peaks.
It is found that the energy gap can controls the amplitude and  width of these resonances and affect their location in the density of states profile.

\vspace{3cm}


\end{abstract}
\end{center}
\end{titlepage}

\section{Introduction}
Graphene has been of intense theoretical and experimental interest due to its unusual electronic properties \cite{Novoselov04(art1)}. With a single sheet of graphene being a zero-gap semiconductor much effort was directed toward engineering gap in the electronic spectrum of graphene by controlling its lateral size and shape. Close to Dirac points, the electrons can be effectively modeled by a massless Dirac equation, showing that electrons behave as chiral particles \cite{Castro09n}. The characteristic feature of the massless Dirac electrons in graphene and their linear energy dispersion 
are at the origin of its unique electronic properties that could be of great importance in nanoelectronic applications \cite{Zhang05,Chung10}.
The electronic band structure of graphene 
 involves two nodal zero-gap points $(K, K')$, called Dirac points, in the first Brillouin zone at which the conduction and valence bands touch. This leads to a number of its unusual peculiar electronic properties such as its high electric conductivity \cite{Geim09, Geim07}.

However, there is general argument against the possibility to confine electrons electrostatically in graphene due to Klein tunneling, which hindered the possibility to use this marvelous material in electronic switching devices that require a gate control over charge carriers \cite{Katsnelson06}. Thus
pristine graphene quantum dots (GQDs) will allow electron to escape from any confining electrostatic potential and will not allow for quantum bound states in an electrostatically confined quantum dot. A large amount of research effort were deployed to create a band gap that allows for charge confinement in GQDs in various ways. This feature along with the characteristic size of the quantum dot, usually in the size range of 1-10 nm,  will enable us to control a wide range of applications including highly tunable physicochemical and fluorescence properties as well as other electrical and optoelectronic properties.
The recent advances in controlled manufacturing of high quality GQDs as well as its strictly two-dimensional nature have established graphene as an exceptional candidate for future nano electronic devices \cite{Geim09,Beenakker08,Peres10}. For this reason the experimental activity aimed at confining electrons in GQDs \cite{Bunch05} had an upsurge in recent years.
On the other hand, the application of a magnetic flux in GQDs allows to control and strengthen the possibility of electrostatically confining fermions \cite{Julia13}. It was shown that the magnetic flux shifts the kinematic angular momentum to integer values, hence allowing for states that cannot be confined by electrostatic gate potentials alone \cite{Bardarson09}.

Theoretically, in the absence of a spectral gap, it has been shown that an electrostatically confined QD can only accommodate quasi-linked states
\cite{Matulis08}. At Dirac point, i.e. 
at energy $E=0$, where the valence and  conduction bands touch, electronic transport through QDs of certain shapes has also been considered \cite{Bardarson09}. In this particular situation, strong resonances in the two-terminal conductance have been predicted. However, in the presence of a spectral gap, real bound states have been obtained \cite{Trauzettel07,Recher09}. The physical methods used to  open a gap in the energy spectrum of graphene are of vital importance for future potential applications \cite{Geim09}.

We  study the electrostatic confinement of electrons in a quantum dot of gapped graphene, surrounded by a sheet of undoped graphene, in the presence of the magnetic flux. We assume that the quantum dot edge smearing is much less than the Fermi wavelength of the electrons and much larger than the graphene lattice constant to ensure the validity of our continuum model \cite{Martin14}.
Solving Dirac equation in each region and applying the continuity of our spinors at the boundary enables us to determine the solutions in each region of space and the corresponding energy spectrum. Subsequently, we use the asymptotic behavior of Hankel functions for large arguments to study approximately the density of states (DOS) as a function of magnetic flux $\phi$, energy gap $\Delta$ and applied electrostatic potential $V$. We numerically compute the DOS under suitable selections of the physical parameters and investigate the different oscillatory behaviors and resonances as well as the dependence of the DOS peaks on the quantum momentum numbers.

The manuscript is organized as follows. In section 2, we set our  theoretical model describing electrostatically confined Dirac fermions. The energy spectrum is given in each region of our system. In section 3, we introduce the scattering matrix formalism to determine the DOS in terms of various  physical parameters. We  compute the DOS and present our results, which reflect the effect of magnetic flux and energy gap on the resonant peaks in the DOS. In section 4, we further discuss different numerical results related to the density of the states and conclude our results in the final section.

\section{Theoretical model}
We consider a quantum dot (QD) defined by a gate with finite-carrier density and surrounded by a sheet of undoped graphene, which is connected to a metallic contact in the form of a ring as depicted in Figure 1.

\begin{figure}[H]\centering
\includegraphics[width=6cm,height=5cm]{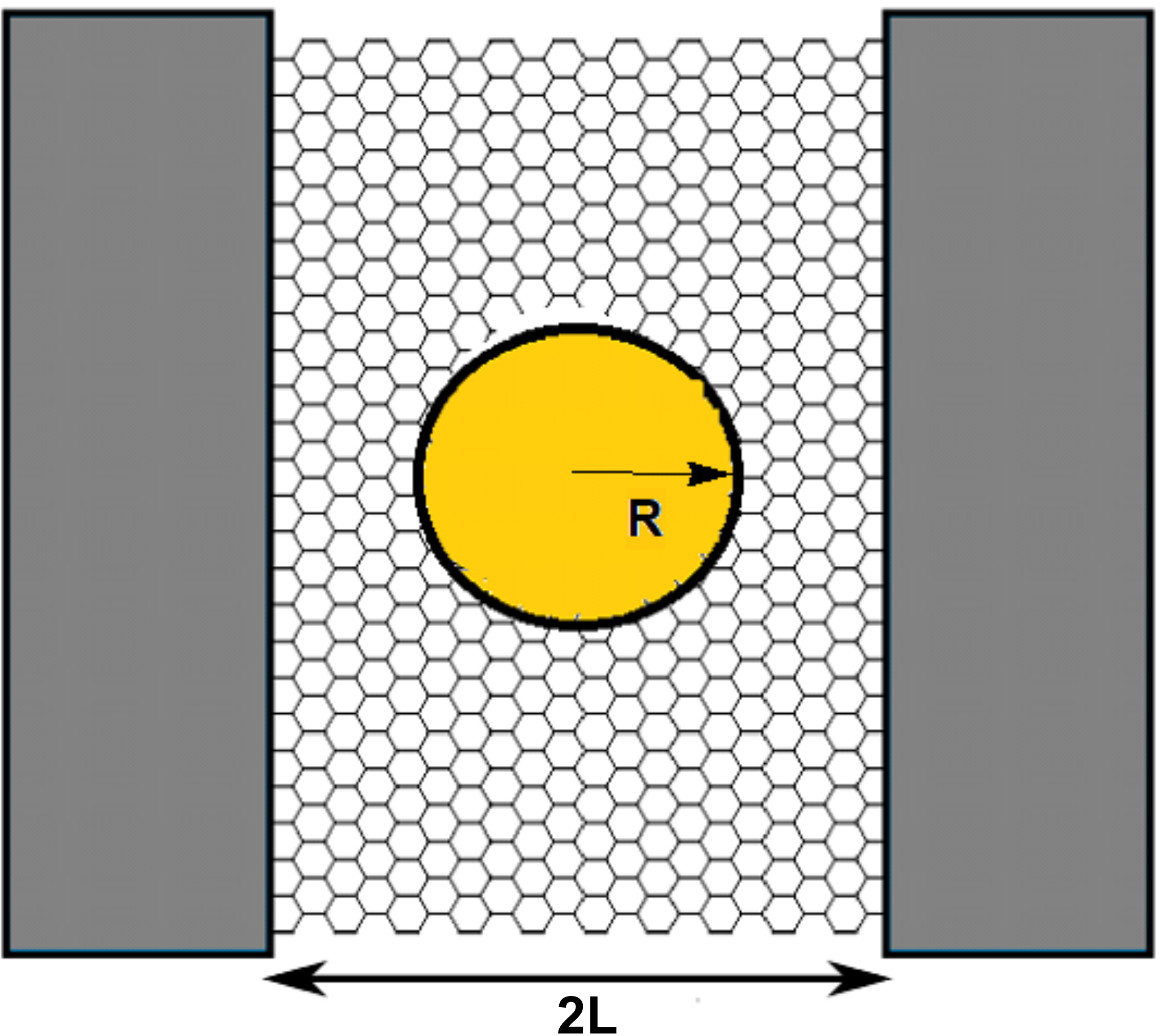}
\caption{\sf (color online) Gate-defined graphene quantum dot (gold color) surrounded by an intrinsic graphene sheet and coupled to a source and drain reservoirs (gray color).  \label{f1}}
\end{figure}
For a Dirac electron in a circular electrostatically defined quantum dot in gapped graphene, the single-valley Hamiltonian can be written as
\begin{equation}
 \label{eq:Dirac}
  H=v_F (\vec p+e\vec{A})\cdot\vec \sigma+V(r) \mathbb{I}+\Delta\sigma_z
\end{equation}
such that the potential barrier $V(r)$ and energy gap $\Delta(r)$ are defined by
\begin{equation}\label{e2}
	V(r)=
	\left\{%
	\begin{array}{ll}
		-\hbar v_F V_0, & r<R\\
         - \hbar v_F V_\infty, & r>L \\
         0,  & \mbox{elsewhere} \\
	\end{array}%
	\right., \qquad
	\Delta(r)=
	\left\{%
	\begin{array}{ll}
	\Delta,  & r< R \\
	0, &   \mbox{elsewhere} \\
	\end{array}%
	\right.
\end{equation}
where $v_F = 10^6$ m/s is the Fermi velocity, $p=(p_x,p_y)$ is the  momentum operator,
$\sigma_i$ are Pauli  matrices in the basis of the two sublattices of $A$ and $B$ atoms.
We choose the parameters $V_0$ and $V_{\infty}$ to be positive, such that dot and lead region are electron-doped. The metallic contact for $r > L$ is modeled by taking the limit $V_{\infty} \to \infty$. The chief reason for our choice of a piece-wise uniform potential is to simplify our analytic  calculations.

In the polar coordinate system $(r,\theta)$,  we introduce the vector potential that generates a solenoid type of magnetic flux
$
\vec{A}(r)=\frac{\hbar}{e}\frac{\phi}{r}\vec{e}_\theta
$
so that the magnetic flux $\phi$ is measured in units of flux quantum $h/e$ and $\vec{e}_\theta$ is the unit vector along the azimuthal direction.
Now the Hamiltonian \eqref{eq:Dirac} takes the form
\begin{equation}
  \label{eq:Hpolar}
  H = \left(\begin{array}{cc} V(r)+\Delta&D_-\\D_+&V(r)-\Delta\\\end{array}\right)
\end{equation}
where the ladder operators are given by
\begin{equation}
 D_{\pm}= -i \hbar v_F e^{\pm i \theta} \left(\partial_r \pm i \frac{1}{r} \partial_{\theta} \mp \frac{\phi}{r} \right).
\end{equation}
Knowing that the total angular momentum $J_z=L_z+\hbar\sigma_z/2$ commutes with the Hamiltonian \eqref{eq:Dirac}, then we look for eigenspinors that are common eigenvectors of both $H$ and  $J_z$. These are
\begin{equation}\label{e2}
\Psi(r,\theta)=
e^{im\theta}\left(%
\begin{array}{c}
  e^{-i\theta/2}\chi_1(r) \\
  i e^{i\theta/2}\chi_2(r) \\
\end{array}%
\right)
\end{equation}
where $m=\pm1/2,\pm3/2 \cdots $ are eigenvalues  of $J_z$.

In the forthcoming analysis, we solve the Dirac equation $H \Psi = E \Psi$ in the three regions: $0 < r < R$, $R < r < L$ and $r > L$.
We obtain
\begin{equation}\label{e8}
\left(\frac{\partial}{\partial r}+\frac{1}{r} \left((m+\phi)+\frac{1}{2}\right)\right)\chi_2(r)={(\epsilon-V_i-\delta)\chi_1}(r)
\end{equation}
\begin{equation}\label{e88}
\left(-\frac{\partial}{\partial r}+\frac{1}{r} \left((m+\phi)-\frac{1}{2}\right)\right)\chi_1(r)={(\epsilon-V_i+\delta)\chi_2}(r)
\end{equation}
 where the dimensionless parameters are used $\epsilon=\frac{E}{\hbar v_F}$, $V_i=\frac{V}{\hbar v_F}$, $\delta=\frac{\Delta}{\hbar v_F}$. For region $R<r<L$ and when $\epsilon=0$, the radial components have the forms
\begin{equation}
\label{psinoenergy}
\chi_1(r)= a_{+} r^{m+\phi-\frac{1}{2}}, \qquad \chi_2(r)= a_{-} r^{-m-\phi-\frac{1}{2}}.
\end{equation}
and to avoid divergence, we impose constraints to fulfill this requirement, $a_{+}=0$ for $m>0$ and $a_{-}=0$ for $m<0$.

Now we consider our system in the absence of magnetic flux $\phi=0$. Then \eqref{e8} and \eqref{e88} reduce to the following equations
\begin{equation}\label{e8b}
\left[\frac{\partial}{\partial r}+\frac{1}{r} \left((m+\frac{1}{2}\right)\right]\chi_2(r)={(\epsilon-V_i-\delta)\chi_1}(r)
\end{equation}
\begin{equation}\label{e88b}
\left[-\frac{\partial}{\partial r}+\frac{1}{r} \left((m-\frac{1}{2}\right)\right]\chi_1(r)={(\epsilon-V_i+\delta)\chi_2}(r).
\end{equation}
Injecting  \eqref{e8b} into \eqref{e88b} to get a second order differential equation for $\chi_1(\rho)$
\begin{equation}\label{e9}
\left[\rho^2 \frac{\partial^2}{\partial \rho^2}+\rho
\frac{\partial}{\partial \rho}+  \rho^2 - \left(m-\frac{1}{2}\right)^2
\right]\chi_1(\rho)=0
\end{equation}
where we have set the variable $\rho=\kappa r$ and the wave number $\kappa$  is defined,
according to each region,  by
\begin{equation}\label{kappa}
 \kappa =\begin{cases}
       \kappa_0=  \sqrt{|(\epsilon+V_0)^2-\delta^2|}, & r<R\\
          \kappa=\epsilon, & R < r < L \\
         \kappa_{\infty}=\epsilon+V_\infty,  &r > L \\
        \end{cases}
\end{equation}
\eqref{e9} has the Hankel function of first $H^{+}_n\left(\rho\right)$ and second $H^{-}_n\left(\rho\right)$ kinds as solutions. Then, we combine all to end up with the eignespinors
\begin{equation} \label{eq:psiref}
\psi_{\kappa,m}^{\pm}(r) =e^{i m \theta} \sqrt{\frac{\kappa}{4\pi}}\begin{pmatrix}
e^{-i\theta/2 }H^\pm_{|m|-1/2}(\kappa r) \\
i~\mathrm{sign}(m)e^{i\theta/2}H^{\pm}_{|m|+1/2}(\kappa r)
\end{pmatrix}
\end{equation}
With the requirement that the wave function is regular at $r=0$, we have the solution inside the quantum dot $r<R$
\begin{equation} \label{eq:psidot}
\psi_{\kappa,m}(r) =e^{i m \theta} \sqrt{\frac{\kappa}{4\pi}}\begin{pmatrix}
e^{-i\theta/2 }J_{|m|-1/2}(\kappa r) \\
i~\mathrm{sign}(m)e^{i\theta/2}J_{|m|+1/2}(\kappa r).
\end{pmatrix}
\end{equation}
Note that the Hankel functions are related to the Bessel $J_n$ and Neumann  $Y_n$ functions by the relations
$
H_n^{(\pm)}=J_n\pm iY_n.$

The presence of the flux $\phi=1/2$ modifies the eigenspinors \eqref{eq:psiref}, because the kinematic angular momentum will be replaced by the canonical one, i.e. $J_{z,kin}=J_z+\hbar \sigma_z/{2}$. We label the new basis states by the integer indices $\mu= m+1/2$ that are eigenvalues of $J_{z,kin}$.
For nonzero $\mu$, the eigenspinors now read as
 \begin{equation} \label{eq:psirefflux}
\psi_{\kappa,\mu}^{\pm}(r) =\sqrt{\frac{\kappa}{4\pi}}\begin{pmatrix}
e^{i (\mu-1) \theta }H^\pm_{|\mu|-1/2}(\kappa r) \\
i~\mathrm{sign}(\mu)e^{i\mu \theta}H^{\pm}_{|\mu|+1/2}(\kappa r)
\end{pmatrix}
\end{equation}
 Note that the half-integer Bessel functions are $Y_{1/2}(x)= - J_{-1/2}(x)= -\sqrt{\frac{2}{\pi x}} \cos x$, $Y_{-1/2}(x)=J_{1/2}(x)=\sqrt{\frac{2}{\pi x}}\sin x$ and $\frac{\cos(kr)}{\sqrt{r}}$ diverges  at the origin.
 Then for $\mu=0$, we have the eigenspinors
 \begin{equation} \label{eq:psirefflux0}
\psi_{\kappa,0}^{\pm}(r) =\frac{e^{\pm i \kappa r}}{\sqrt{8 \pi^2 r}}\begin{pmatrix}
\pm e^{- i \theta } \\
1
\end{pmatrix}.
\end{equation}

In the next, we will show how the above results can be used to analyze the density of states
associated to our system. In fact, it will be done by distinguishing two cases:
without and with magnetic flux.

\section{Density of states}

To give a better understanding of the basic features of our system, let us
investigate the density of states (DOS). For this, 
we introduce the local DOS $\nu(r,\epsilon)$ 
that is given in terms of the scattering matrix $\mathcal{S}(\varepsilon)$ \cite{Langer1961,Buttiker1993,Buttiker1994}
\begin{equation}
  \nu(r,\epsilon) = \frac{1}{2 \pi i \hbar v_F } \mbox{Tr}\, {\cal S}^{\dagger} \left( \frac{\delta {\cal S}}{\delta V(r)}+\frac{\delta {\cal S}}{\delta \Delta(r)}\right)
\end{equation}
such that $\mathcal{S}(\varepsilon)$ can be determined using the boundary conditions.
Now
to get  the total DOS, we simply integrate over the region $r < L$ 
to end up with  
\begin{equation}
  \nu_{dot}(\epsilon) = \frac{1}{2 \pi i \hbar v_F}
  \int_{r < L} \mbox{Tr}{\cal S}^\dagger \left( \frac{\delta {\cal S}}{\delta V(r)}+\frac{\delta {\cal S}}{\delta \Delta(r)}\right)~dr.
  \label{eq:nuWS}
\end{equation}
To calculate $\nu_{\rm dot}$ at zero energy $(\epsilon=0)$ as a function of the quantum dot parameters, it suffices to solve the Dirac equation associated to the Hamiltonian \eqref{eq:Dirac}, at small but finite energy $\epsilon$, and determine the scattering matrix $S$. This will be done by considering the zero and nonzero magnetic flux cases.
 
\subsection{Zero magnetic flux}
In the present case and for $r > L$, the eigenspinors can be written as a linear combination of the two solutions of \eqref{eq:psiref}
\begin{equation}
 \label{eq:psi}
  \psi_{\epsilon,m}(r)= a_{m}(\epsilon) \psi^{-}_{k_\infty,m} (r)+b_{m}(\epsilon) \psi^{+}_{k_\infty,m} (r).
\end{equation}
To determine the coefficients $a_{m}(\epsilon)$ and $b_{m}(\epsilon)$, we use the boundary conditions at interfaces $r=L$ and $r=R$, together with the regularity at $r=0$. This process allows to obtain
\begin{equation}
  \label{eq:scatt}
  b_{m}(\epsilon) = \mathcal{S}_{m}(\epsilon) a_{m}(\epsilon)
\end{equation}
such that the scattering matrix $\mathcal{S}_{m}(\epsilon)$ reads as
\begin{equation}
\label{S}
S_{m}(\epsilon) = - \frac{\det D^{(-)}}{\det D^{(+)}}
\end{equation}
where both matrices are given by
\begin{equation}
\label{D12}
D^{(+,-)}=
\begin{pmatrix}
 0 & \sqrt{\kappa} H^{(+)}_{|m|-\frac{1}{2}} (\kappa R) &\sqrt{\kappa} H^{(-)}_{|m|-\frac{1}{2}} (\kappa R) & \sqrt{\kappa_\infty} J_{|m|-\frac{1}{2}}(\kappa R) \\
0 & \sqrt{\kappa} H^{(+)}_{|m|+\frac{1}{2}} (\kappa R) & \sqrt{\kappa} H^{(-)}_{|m|+\frac{1}{2}} (\kappa R) & \sqrt{\kappa_\infty} J_{|m|+\frac{1}{2}}(\kappa_0 R) \\
\sqrt{\kappa_0} H^{(-,+)}_{|m|-\frac{1}{2}} (\kappa_\infty L) & - \sqrt{\kappa} H^{(+)}_{|m|-\frac{1}{2}} (\kappa L) & - \sqrt{\kappa} H^{(-)}_{|m|-\frac{1}{2}} (\kappa L) & 0 \\
\sqrt{\kappa_0} H^{(-,+)}_{|m|+\frac{1}{2}} (\kappa_\infty L) & - \sqrt{\kappa} H^{(+)}_{|m|+\frac{1}{2}} (\kappa L) & - \sqrt{\kappa} H^{(-)}_{|m|+\frac{1}{2}} (\kappa L) & 0
\end{pmatrix}.
\end{equation}
We consider the limit of a highly doped lead $k_{\infty} L\gg 1$ to approximate the asymptotic behavior of the Hankel functions for large arguments as
\begin{equation}
H^{(\pm)}_n(x)\approx (2/\pi x)^{1/2} e^{\pm i(x-n\frac{\pi}{2}-\frac{\pi}{4})}
\end{equation}
which is valid  in the lead region $r > L$.
For a short-distance, we have
\begin{equation}
J_{n}(x)\sim \frac{1}{n!} \left(\frac{x}{2}\right)^{n},
\qquad
Y_{n}(x) \sim \left\{\begin{array}{c} -\frac{\Gamma(n)}{\pi}\left(\frac{2}{x}\right)^n,~~n>0\\ \frac{2}{\pi}\ln\left(\gamma_E\frac{x}{2}\right),~~n=0
\end{array}  \right.
\end{equation}
where $\ln\left(\gamma_E\right)=0.577\cdots$ is the Euler’s constant. For negative integer, $m<0$, we have the relation $J_{-m}=(-1)^m J_m$ and $Y_{-m}=(-1)^m Y_m$. For small energy $\epsilon$, we can develop the scattering matrix as a function of $\kappa$ in the region $R <r <L$ and choose $\chi_{1,2}(r)\propto J_{n}(kr)$ regular at $r=0$. We then find
\begin{align}
  \mathcal{S}_{m}(\epsilon) = e^{-2 i \kappa_{\infty} L + i |m| \pi}
  \left[ \mathcal{S}_{m}^{(0)} + \kappa \mathcal{S}_{m}^{(1 )}
  + {\cal O}(\epsilon^2) \right]
\end{align}
such that
\begin{equation}
 \label{eq:calS}
 \mathcal{S}^{(0)}_{m}=\frac{L^{2|m|}+i {\cal J}_m R^{2|m|}}
  {L^{2|m|} -i {\cal J}_m R^{2|m|}}
\end{equation}
and $S_{m}^{1}$ takes the following forms for $m\neq \frac{1}{2}$
\begin{equation}
 \label{eq:calS1m}
 \mathcal{S}^{(1)}_{m}=
  \displaystyle
  -\frac{2 i L}{2|m|-1} {\cal S}^{(0)}_m
  +
  \frac{8 i |m| L^{4 |m|+1} +2 i[(2|m|+1){\cal J}_m^2-(2|m|-1)]R^{2|m|+1} L^{2|m|}}{(4 |m|^2 - 1)(L^{2 |m|} - i {\cal J}_m R^{2|m|})^2}
\end{equation}
or for $m=\pm \frac{1}{2}$
\begin{equation}
\label{eq:S1m12}
 \mathcal{S}^{(1)}_{\pm 1/2}=
  \frac{i L (L^2-R^2)+2 i{\cal J}_{\frac{1}{2}}^2 R^2 L \ln (L/R)}{(L-i{\cal J}_{\frac{1}{2}}R)^2}
\end{equation}
where ${\cal J}_m$ is given by
\begin{equation}
 {\cal J}_m=\frac{J_{|m|+1/2}(\kappa_0 R)}{J_{|m|-1/2}(\kappa_0 R)}.
\end{equation}
We now use \eqref{eq:nuWS} to calculate DOS $\nu_{\rm dot}$ at zero energy for the both cases $m\neq \frac{1}{2}$,  $m= \frac{1}{2}$. Then, our calculation shows 
\begin{equation}
 \label{eq:deltanu}
 \nu_{\rm dot} = \frac{1}{2 \pi i \hbar v_F} \sum_{m}
  ~ \mathcal{S}^{(0)*}_m \left[ \frac{\partial \mathcal{S}^{(0)}_m}{\partial \kappa_0}
  + \mathcal{S}^{(1)}_m \right] \left[\frac{\partial \kappa_0}{\partial V_0}-\frac{\partial \kappa_0}{\partial \delta}\right].
\end{equation}
The first term in \eqref{eq:deltanu} represents the integral of the local DOS inside the QD region ($r < R$), while the second one its  integral in the undoped layer that separates QD and the metallic contact \cite{Martin14}.
On the other hand, for zero energy and by using  the  continuity of the eigenspinors \eqref{eq:psidot} and \eqref{psinoenergy} at $r = R$, we find the resonance condition
\begin{equation}
\label{eq:resonanceposition}
 J_{|m|-1/2}(\kappa'_0 R)=0
\end{equation}
where $\kappa_0=\kappa'_0$.
In the limit $R \ll L$, DOS exhibits isolated resonances at gate values 
close to resonance, we can then write
 \begin{equation}
{\cal J}_{m}\approx \frac{-1}{R(\kappa_0 - \kappa'_0)}
\end{equation}
showing that DOS has a Lorentzian dependence on $\kappa_0$.
Now
for $|m| \neq 1/2$, the zero-energy DOS takes the form
 \begin{equation}
 \label{eq:deltanures}
  \nu_{dot} = \frac{4 R |m|}{\pi \hbar v_F (2|m|-1)}
  \frac{\Gamma}{4 R^2 (\kappa_0 - \kappa'_0)^2 + \Gamma^2} \frac{|V_0-\delta|}{\kappa_0}
\end{equation}
whereas for $|m| = 1/2$, it reads as
\begin{equation}
  \nu_{\rm dot} = \frac{2 R}{\pi \hbar v_F}
  \left(1 + \ln \frac{L}{R} \right)
  \frac{\Gamma}{4 R^2 (\kappa_0 - \kappa'_0)^2 + \Gamma^2} \frac{|V_0-\delta|}{\kappa_0}
\end{equation}
where we have set the parameter of our theory as $\kappa_0=\sqrt{|V_0^2-\delta^2|}$ and  the dimensionless resonance width is given by
\begin{equation}
  \label{eq:Gamma}
  \Gamma=2\left( \frac{R}{L}\right)^{2|m|}.
\end{equation}

\subsection{Non zero magnetic flux}
We now investigate the density of states for a gapped graphene quantum dot in the presence of the magnetic flux, such that eigenspinors are those in \eqref{eq:psirefflux} and \eqref{eq:psirefflux0} taking into consideration the kinematic angular momentum $\mu$.
The states with zero kinematic angular momentum need to be discussed separately in the presence and absence of magnetic flux. We first discuss the states with $\mu\neq 0$, where magnetic field only leads to slight modifications. Effectively, one finds that the results of  \eqref{S} and \eqref{eq:calS1m} remain valid, as much as the half-integer index $m$ is replaced by the integer index $\mu$. For $\mu \neq 0$, the calculation of  bound states proceeds in the same way as without flux and we find that the resonance condition is given by
 \begin{equation}
 \label{eq40}
 J_{|\mu|-1/2}(\kappa'_0 R)=0.
 \end{equation}
  We conclude that, if the quantum dot and the surrounding undoped graphene layer are contacted to source and drain reservoirs, the width $\Gamma$ of the resonances is
  \begin{equation}
  \label{eq41}
  \Gamma=2(R/L)^{2\mu}.
  \end{equation}
For the case $\mu=0$, regularity of the wave function at the origin is not sufficient to determine the scattering matrix $S_0(\epsilon)$. Taking a flux line of extended diameter, we find the condition of the wave function has to vanish at the origin \cite{heinl2013}. The calculation of the scattering matrix $S_0(\epsilon)$ is straightforward and leads to the following result
\begin{equation}
S_0=e^{-2i(\kappa_{\infty}-\kappa_0)R} e^{-2i(\kappa_{\infty}-\kappa)(L-R)}
\end{equation}
where $\kappa$, $\kappa_0$ and $\kappa_{\infty}$ are given in \eqref{kappa}
as function of the gate voltage $V$ and energy gap $\Delta$. Note that by requiring $\Delta=0$, we recover  both DOS derived in \cite{Martin14}.
The obtained results so far will be numerically analyzed to emphasis the main features
of our system and therefore underline the influence of the energy gap on the quantum dot.

\section{Results and discussions}
We study the influence of the introduced  energy gap $\delta$ and magnetic flux $\phi=1/2$ at energy incident $\epsilon=0$ on the bound states of an electrostatically confined graphene quantum dot of radius $R$ and the contact size  $L$. 
Indeed, because the parameter of our theory is  $\kappa_0=\sqrt{|V_0^2-\delta^2|}$, then we choose to numerically analyze the DOS versus the gate voltage $V_0R$ under suitable conditions
of the physical parameters. 
More precisely, we consider
 particular values of the ratio 
 $R/L$=(0.05, 0.07, 0.1, 0.15, 0.2) 
 energy gap $|\delta R|\leq (0,0.5, 1,2,3,4)$,
angular quantum numbers $m=(1/2,3/2)$ for zero and $\mu=(1,2)$ for nonzero fluxes.\\

\begin{figure}[!hbt]
\centering
	\includegraphics[width=0.49\linewidth]{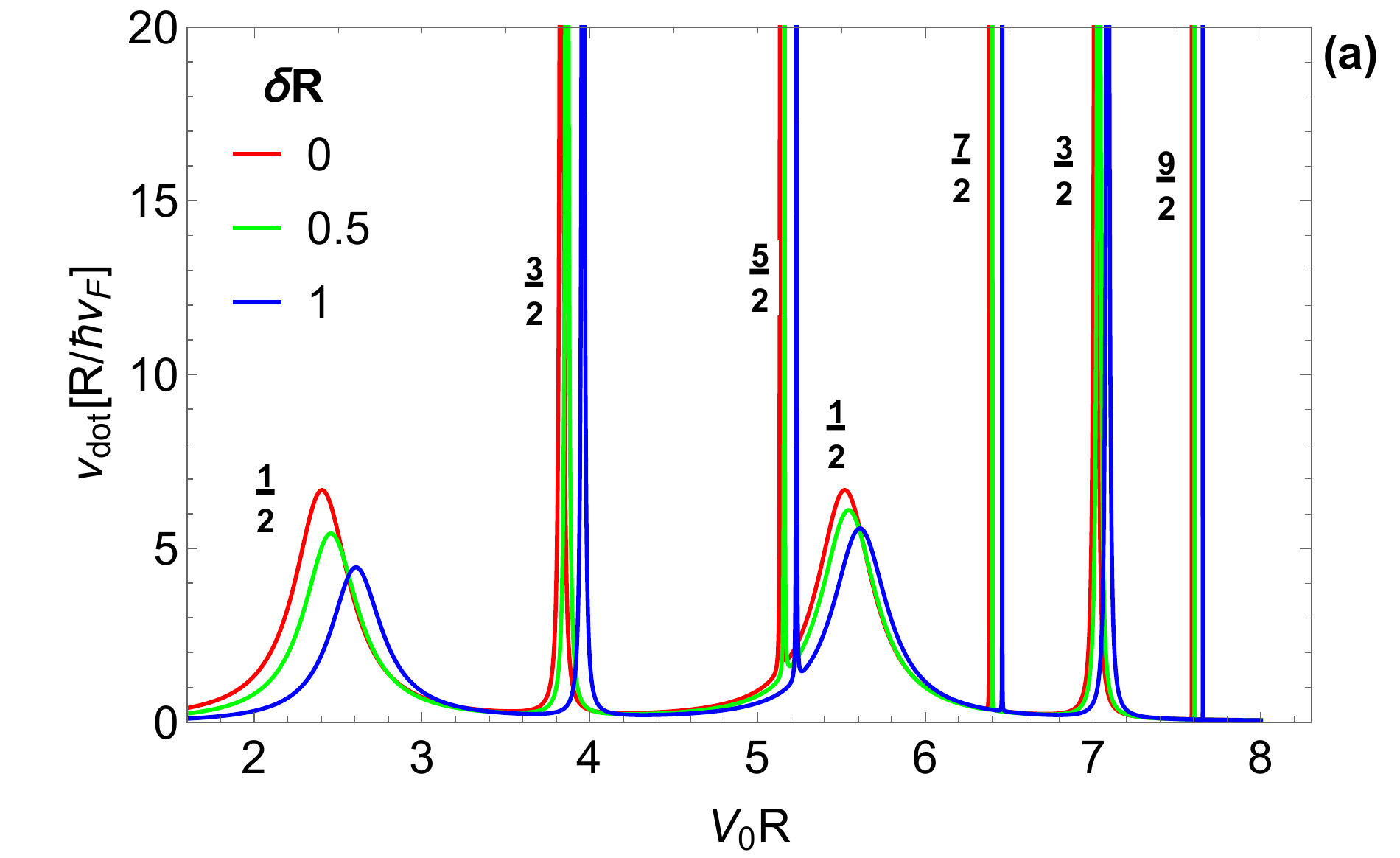}\includegraphics[width=0.49\linewidth]{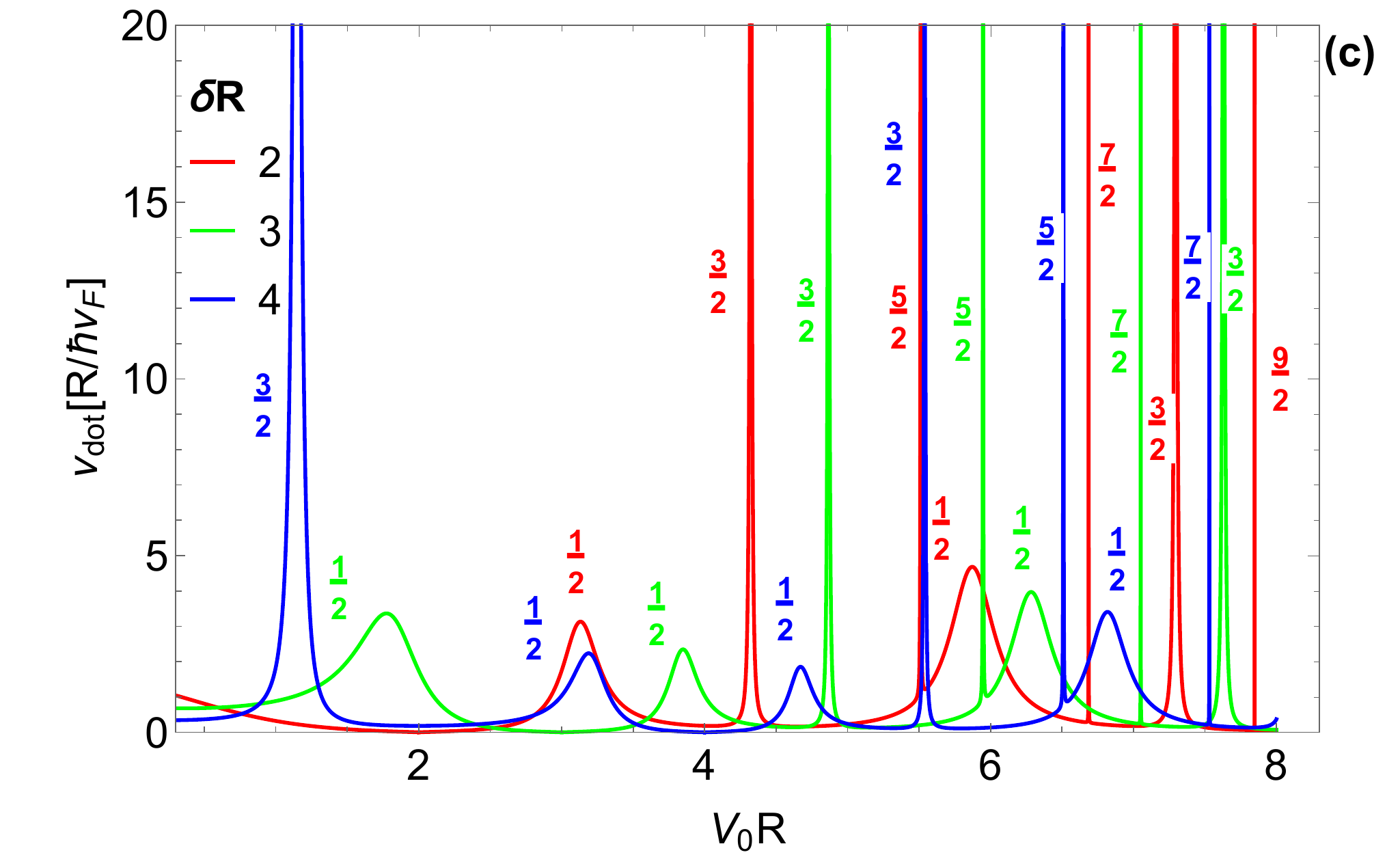}\\\includegraphics[width=0.49\linewidth]{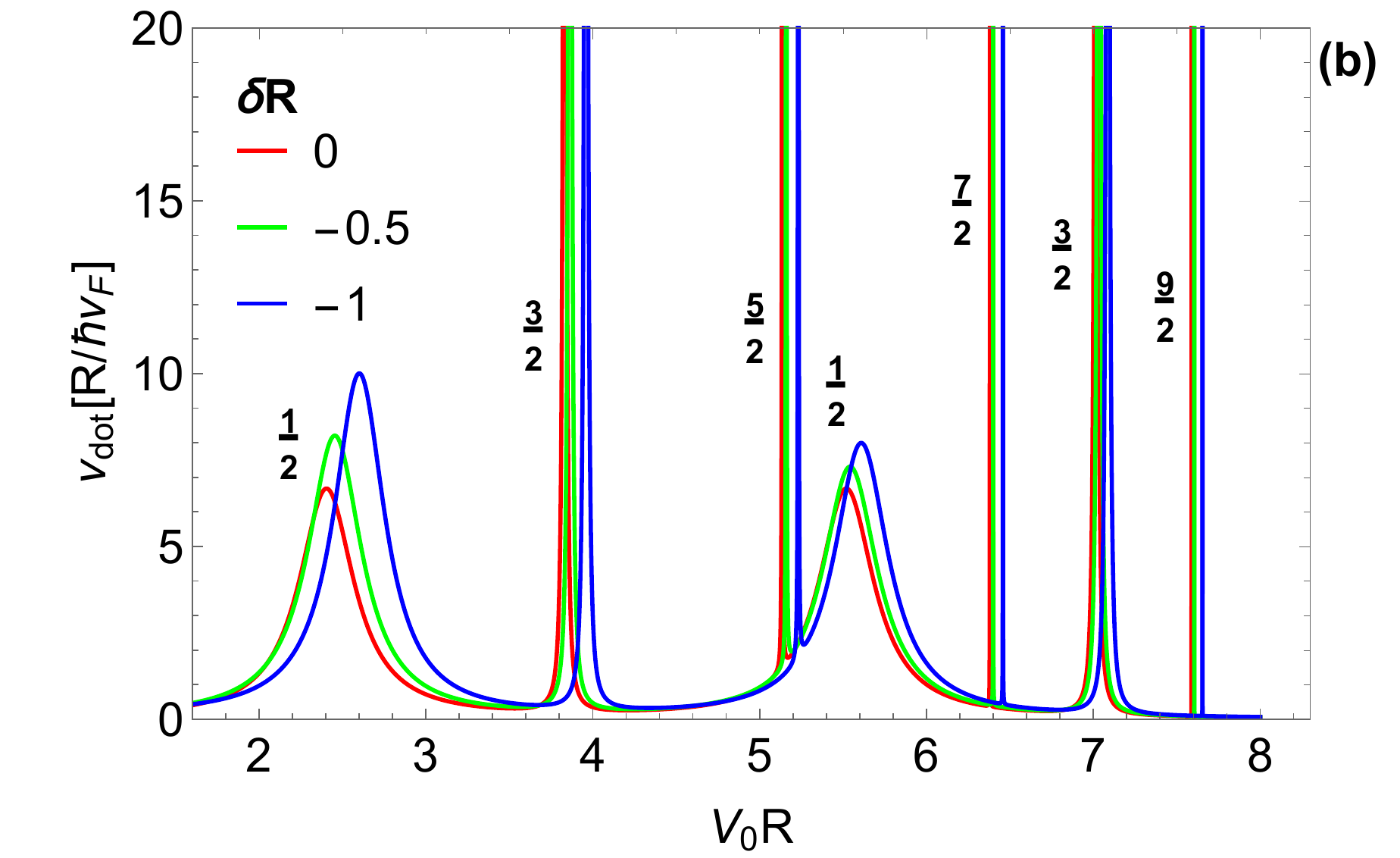}\includegraphics[width=0.49\linewidth]{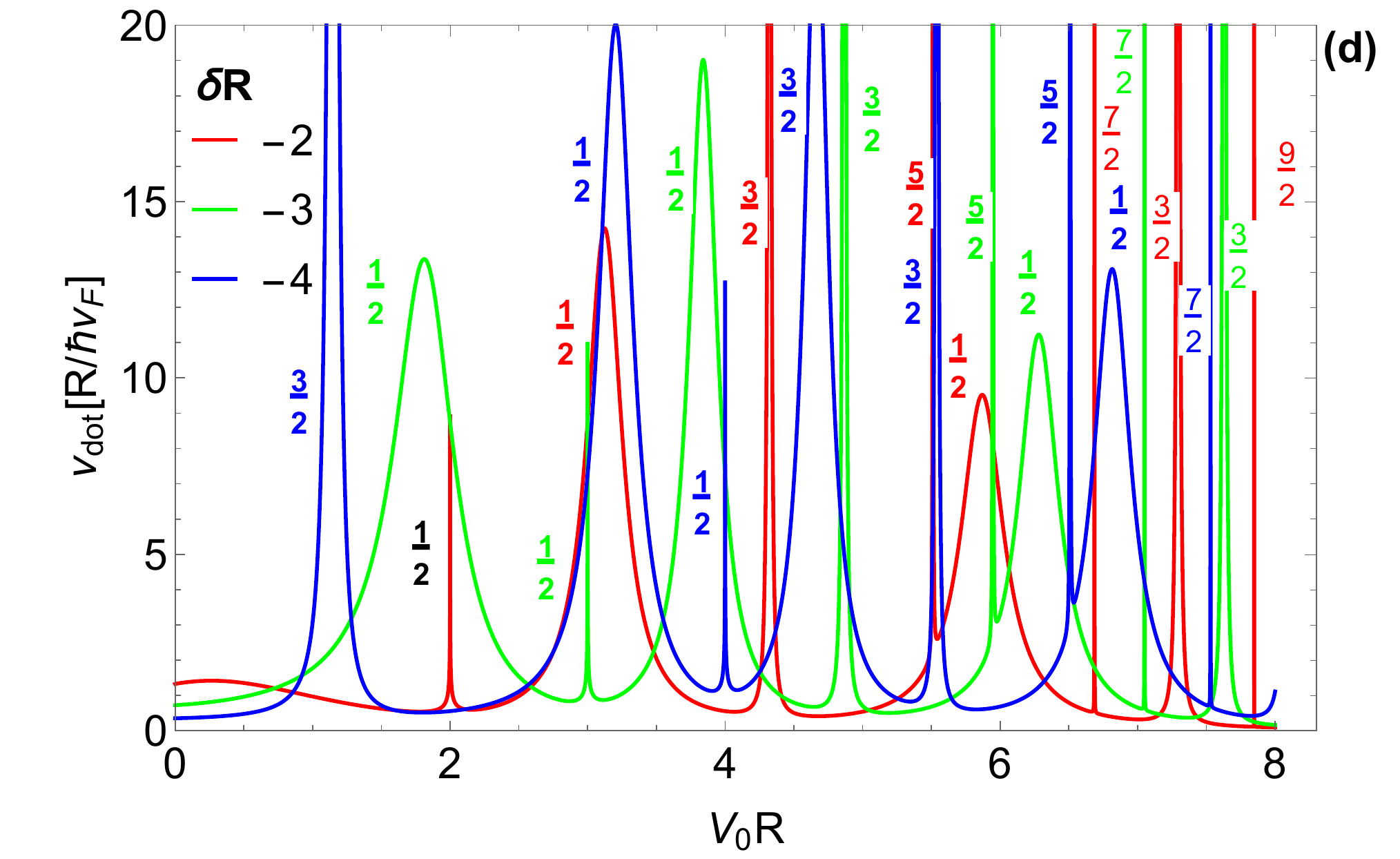}
	\caption{\sf (color online) The DOS as function of the gate voltage  $V_0 R$ at incident energy    $\epsilon =0$ for ratio $R/L = 0.2$ and different values of the energy gap $\delta R$. The resonances are labeled according to their angular momentum $m=\pm 1/2, \cdots, \pm 9/2$. (a): $\delta R=0, 0.5, 1$. (b): $\delta R= 0, -0.5,-1$. (c): $\delta R= 2, 3, 4$. (d): $\delta R=-2, -3, -4$.}
	\label{f2}
\end{figure}

The DOS for a circular quantum dot as function of the gate voltage $V_0R$ at $\epsilon =0$ for  $R/L = 0.2$ and different values of the energy gap $\delta R$, is shown in Figure \ref{f2}. We observe that the DOS exhibits an oscillatory behavior with the appearance of resonance peaks, which are labeled according to their angular quantum  momentum $m$. This behavior shows that when $\delta$ increases the amplitude of DOS decreases with a shift to the right when $\delta R$ is positive see Figure \ref{f2}(a,c). For negative values, the amplitude and  width increase when the absolute value of $\delta$ increases. We also notice that the resonance peaks move towards the left see Figure \ref{f2}(b,d). Note that for $\delta=0$, the position of  resonances as well as  width and  amplitude are in agreement with the results  obtained in the literature \cite{Martin14,Bardarson2009,Titov2010}. It is clearly seen that for higher value of  
$m$, 
the resonance disappear and peaks take places.

\begin{figure}[h]\centering
	\includegraphics[width=0.5\linewidth]{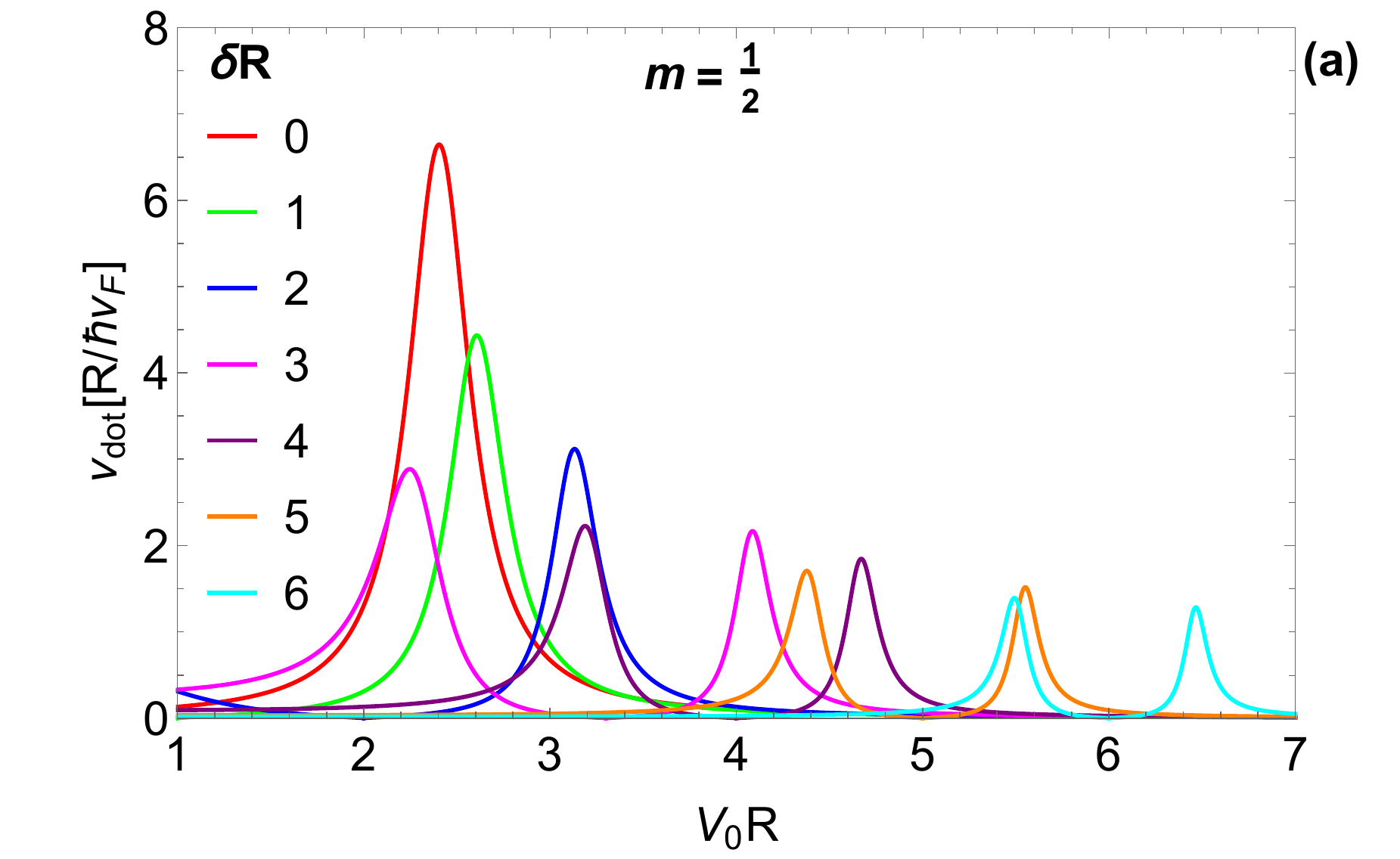}\includegraphics[width=0.5\linewidth]{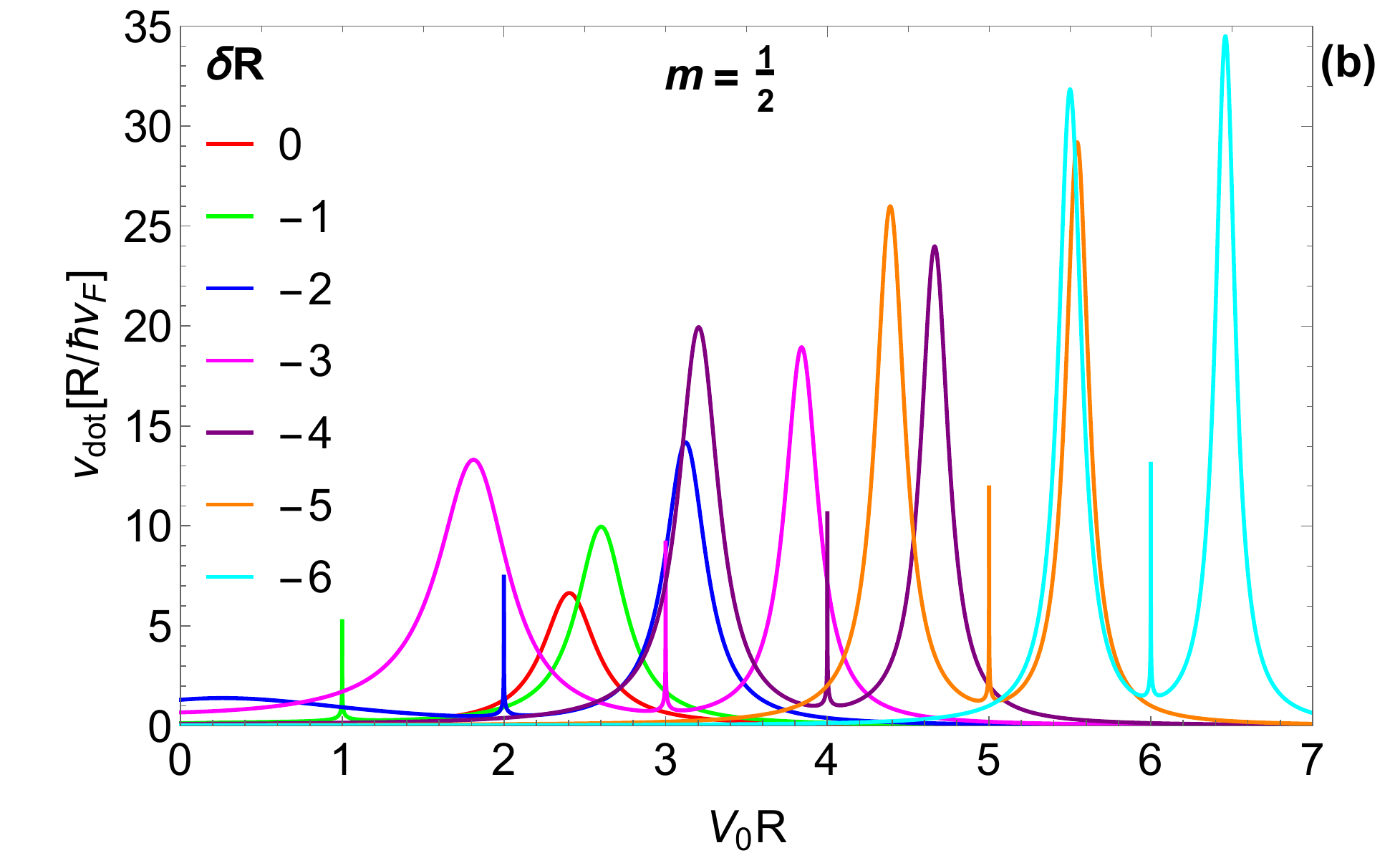}\\
	\includegraphics[width=0.5\linewidth]{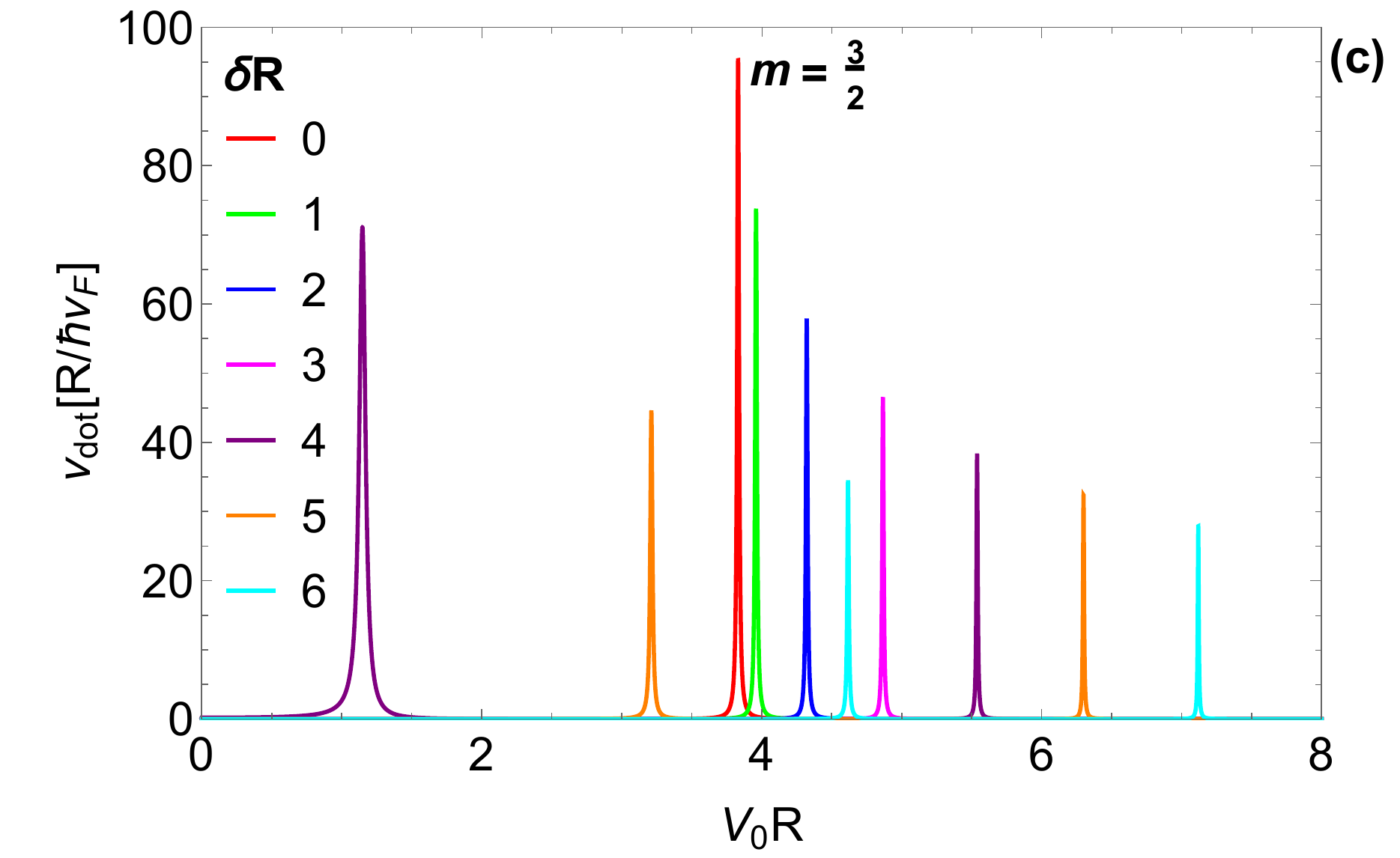}\includegraphics[width=0.5\linewidth]{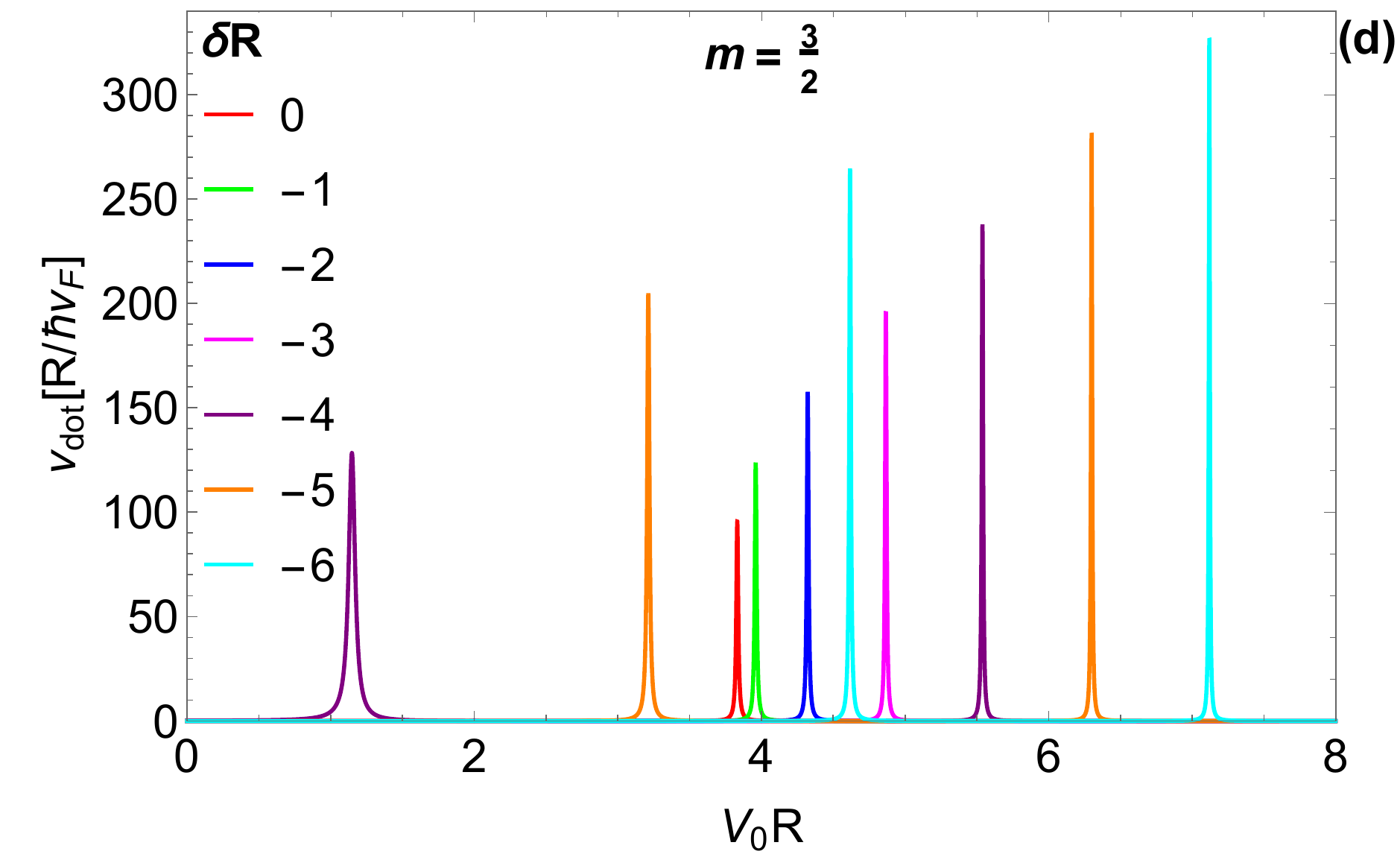}
	\caption{\sf (color online) The DOS as function of the gate voltage $V_{0} R$ incident energy    $\epsilon =0$ for ratio $R/L=0.2$ and different values of the energy gap $\delta R$. (a): $\delta R=0, 1, 2, 3, 4, 5, 6$ and (b): $\delta R=0, -1, -2, -3, -4, -5, -6$ for  first resonance $m=1/2$. (c): $\delta R=0, 1, 2, 3,4, 5, 6$ and (d): $\delta R=0, -1, -2, -3, -4, -5, -6$ for  second resonance $m=3/2$.}\label{f3}
\end{figure}

The plot of the DOS shows clearly the first  resonance $m=1/2$  and second resonance $m=3/2$ for different values of $\delta R=0,\pm1,\pm2,\pm3,\pm4$. We deduce that the resonance characteristics depend on both the sign and magnitude of $\delta$. Indeed, from Figure \ref{f3}(a,c) and for positive $\delta$, the resonance positions shift so that the amplitude and width of the resonance decrease if $\delta R$ increases. Whereas from Figure \ref{f3}(b,d) and for negative $\delta$, there  appear sharp peaks at the location which corresponds to the chosen values of $|\delta R|$. 
We also notice that the first and second resonances are doubled when $\delta R$ exceeds the position $V_0 R\geq |\delta R|=3$ (Figure \ref{f3}(a,c)) and $V_0 R\geq |\delta R|=4$ (Figure \ref{f3} (b,d)). We observe that the DOS exhibits an oscillatory behavior with decreased amplitude when $V_0R$ increases for $\delta\geq 0$ \cite{Martin14} and an increase in the oscillation amplitude for $\delta\leq 0$. Moreover, the width of the second resonance in the DOS  (Figure \ref{f3}(c,d)) is very small compared to first resonance (Figure \ref{f3}(a,b)).

\begin{figure}[h]\centering
	\includegraphics[width=0.33\linewidth]{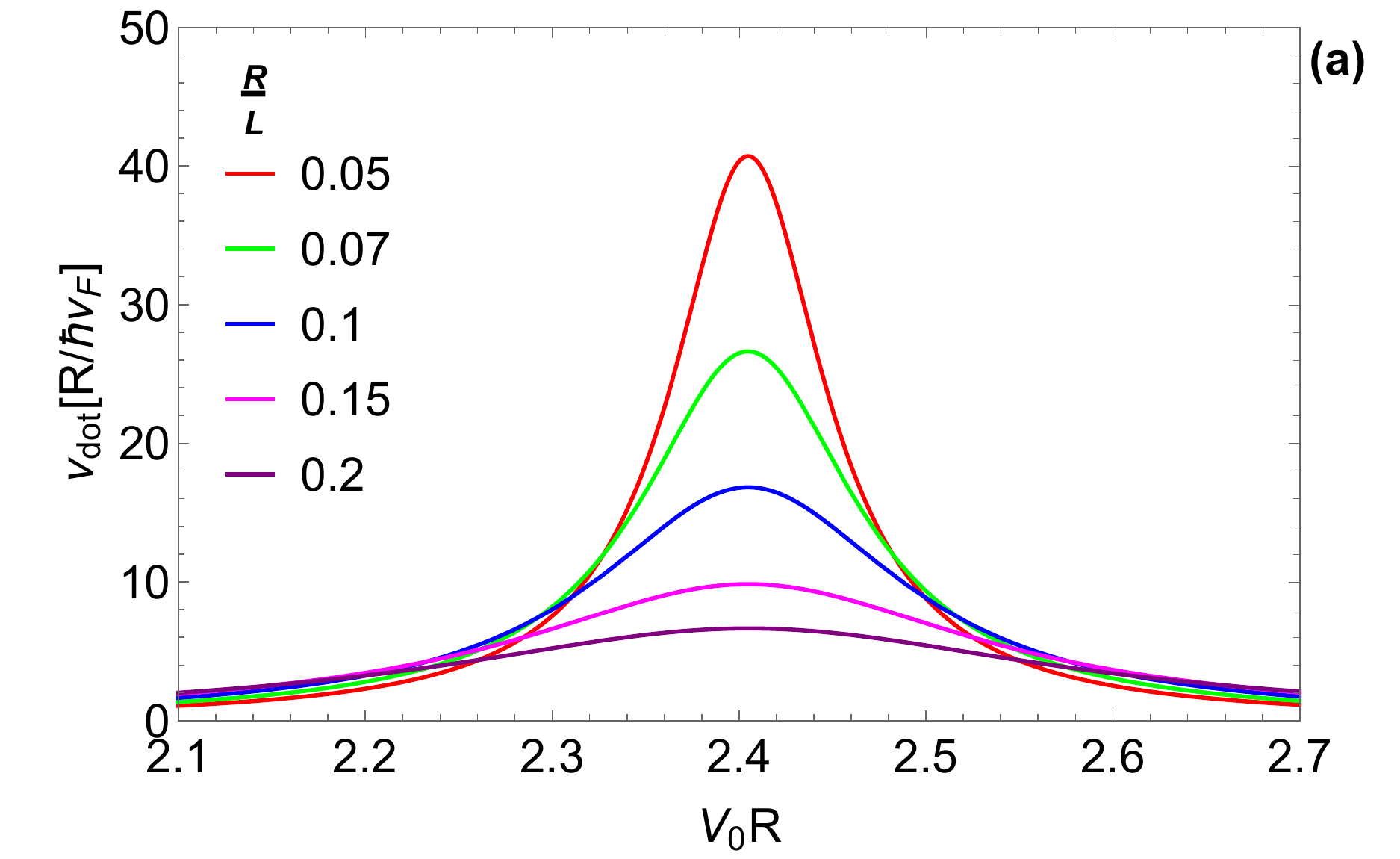}\includegraphics[width=0.33\linewidth]{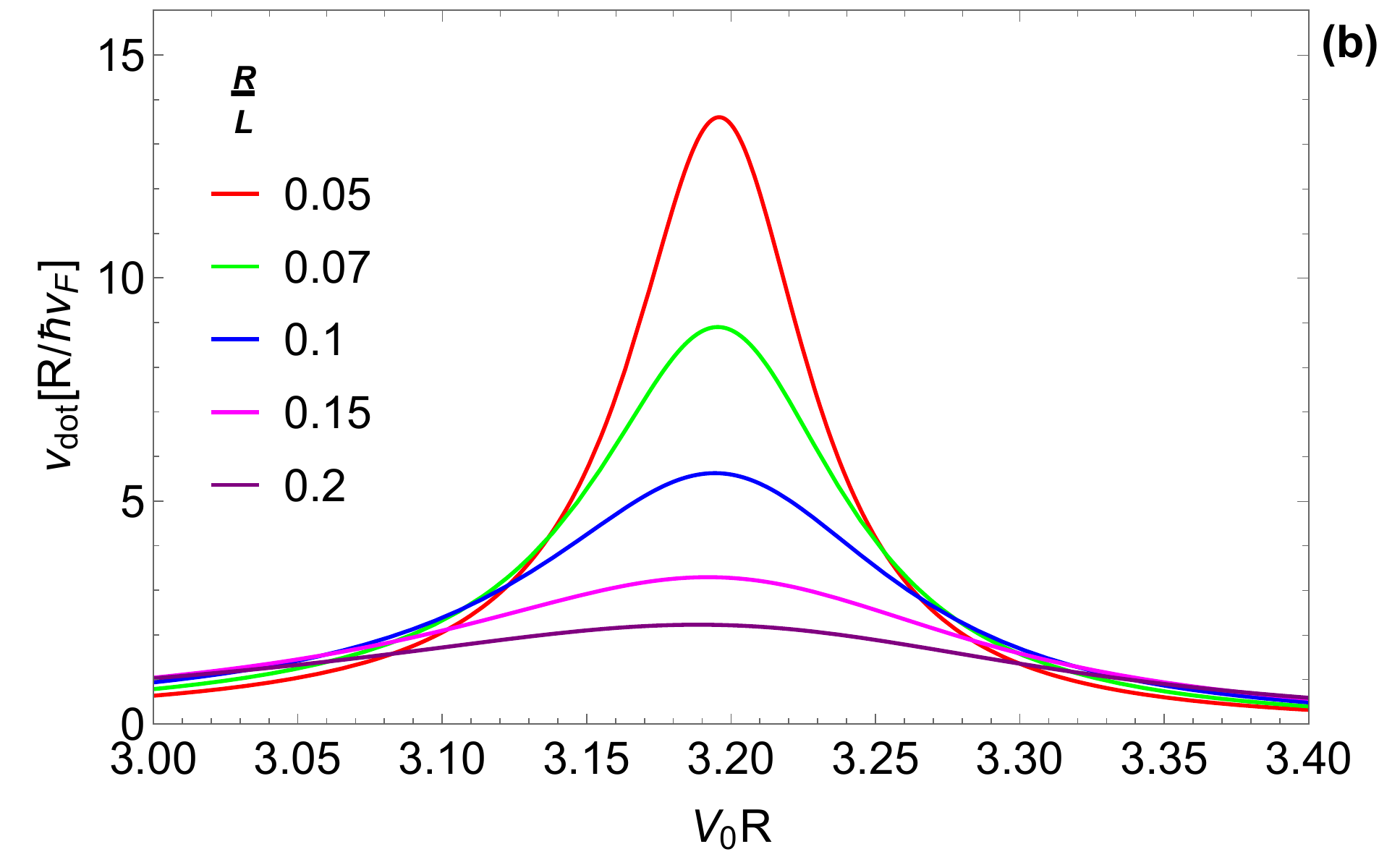}\includegraphics[width=0.33\linewidth]{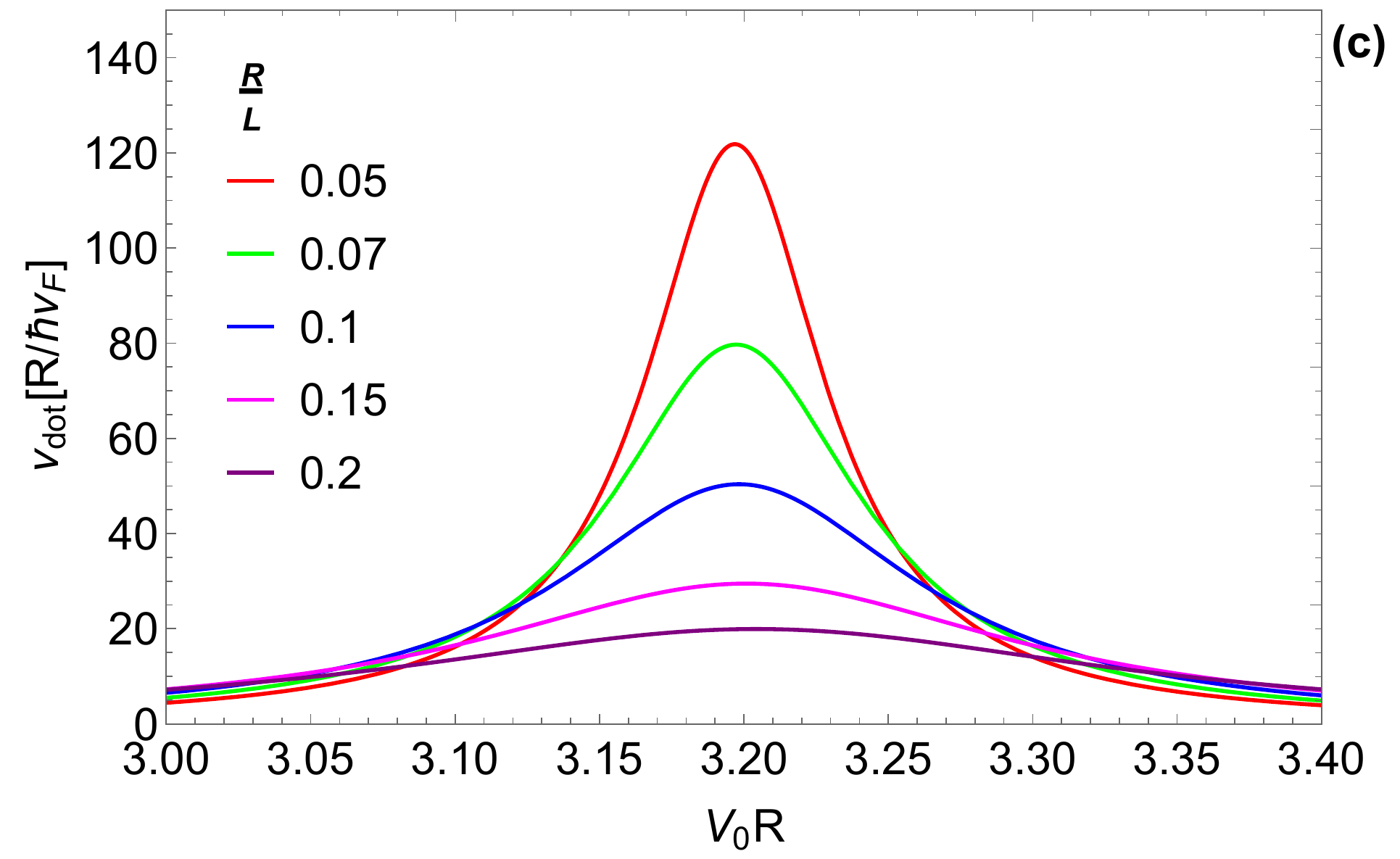}
	\caption{\sf(color online) The  DOS  as function of the gate voltage $V_{0} R$ at incident energy $\epsilon=0$ and first resonance $m=1/2$ for different values of the energy gap and 
		ratio $R/L$. (a):  $\delta R=0$, (b):  $\delta R=4$ and (c): $\delta R=-4$.}
	\label{f4}
\end{figure}

To show how the first resonance $m=1/2$ behaves when we modify the energy gap $\delta R$ and increase the contact size $L$ for a fixed radius $R$, we present the DOS as function of the gate voltage $V_0 R$ in Figure \ref{f4}, with  (a): $\delta R=0$, (b): $\delta R=4$ and (c): $\delta R=-4$.
It is clearly seen that when $R/L$ is very small, the DOS saturates (maximum), which could be explained by invoking the weak coupling between the QD and the metallic contacts \cite{Martin14}. Now by comparing Figures \ref{f4}(a,b,c), we notice that the amplitudes of  resonance become very important for very small and negative value of $\delta R$.\\

\begin{figure}[H]\centering
\includegraphics[width=0.5\linewidth]{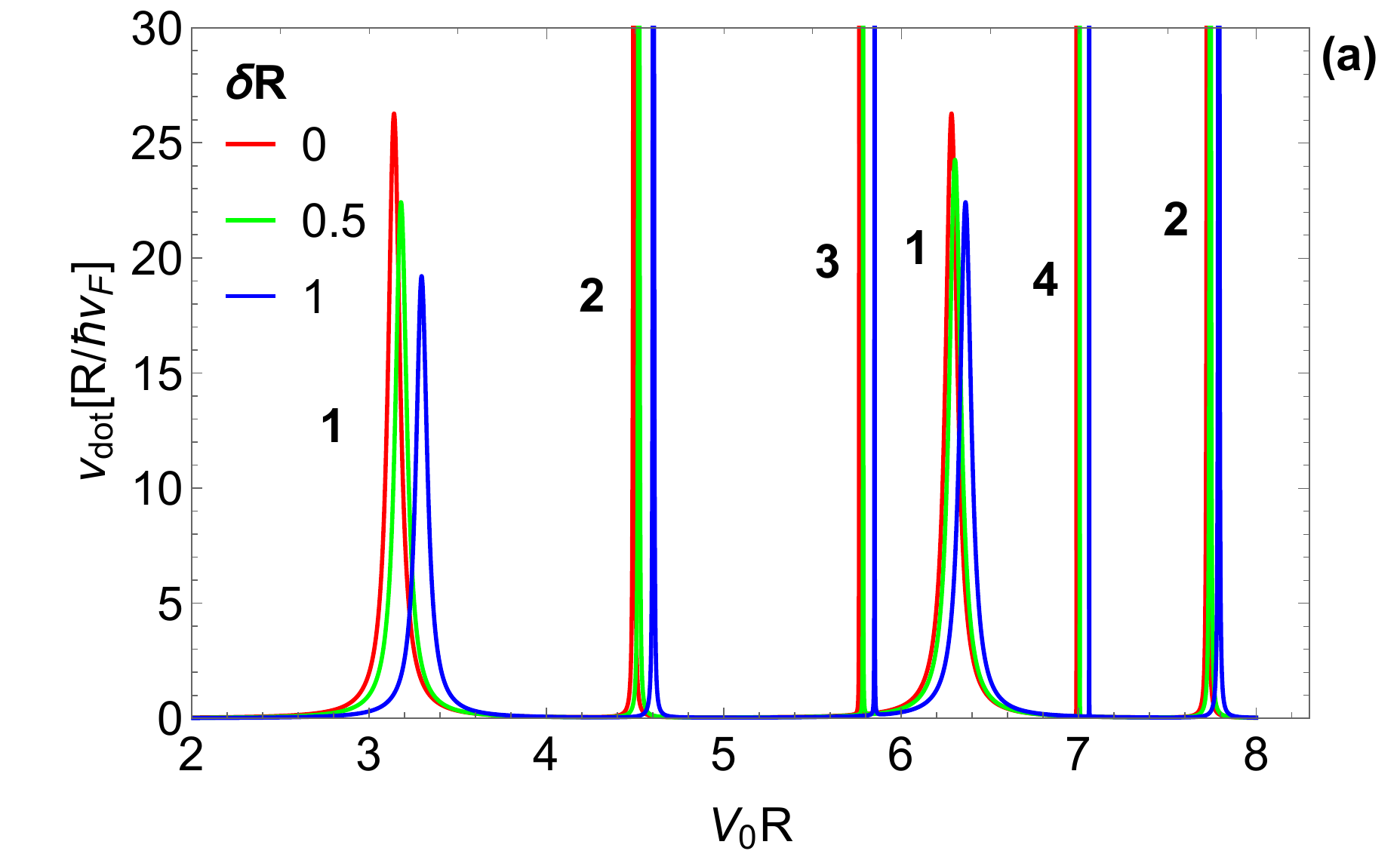}\includegraphics[width=0.5\linewidth]{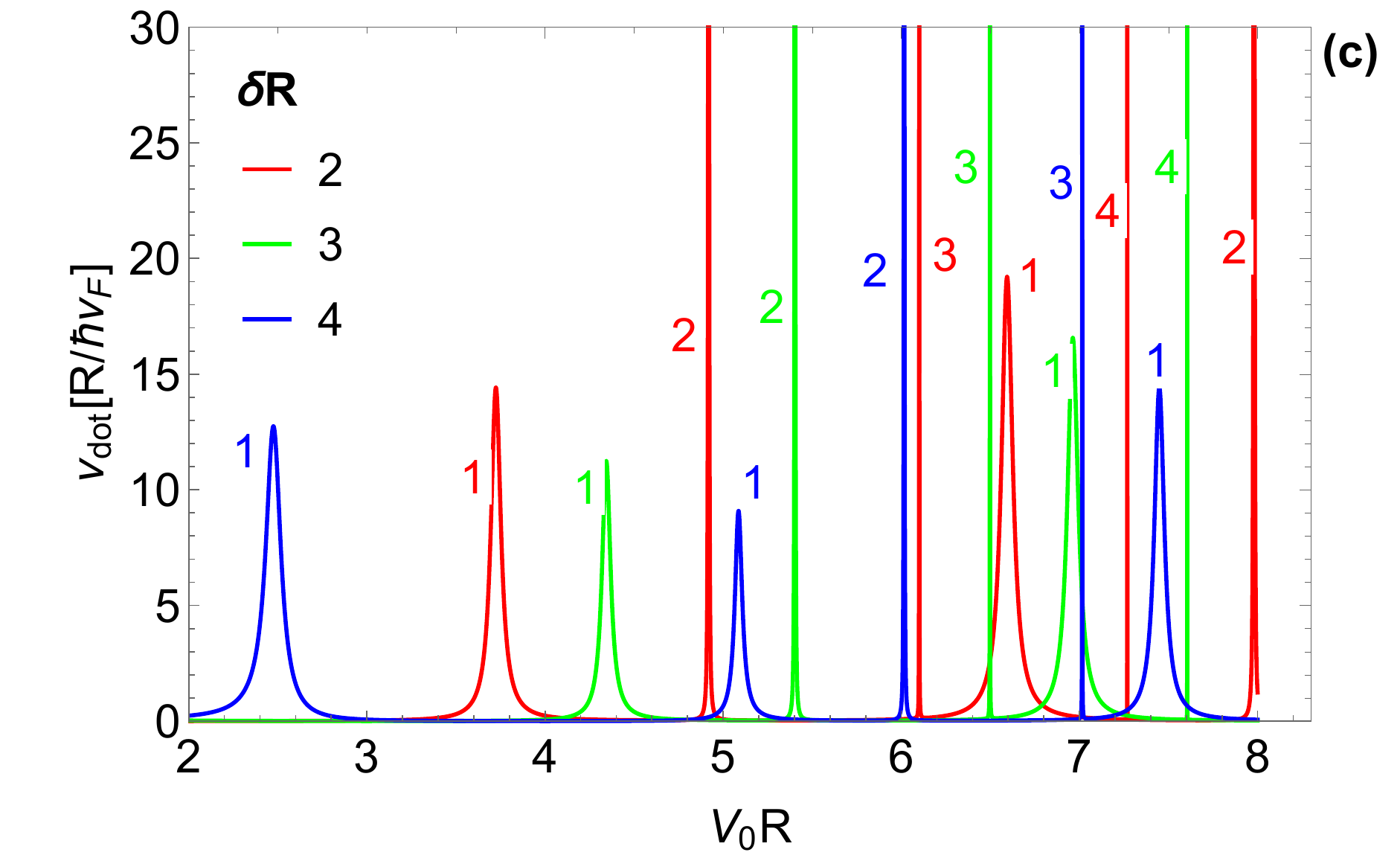}\\\includegraphics[width=0.5\linewidth]{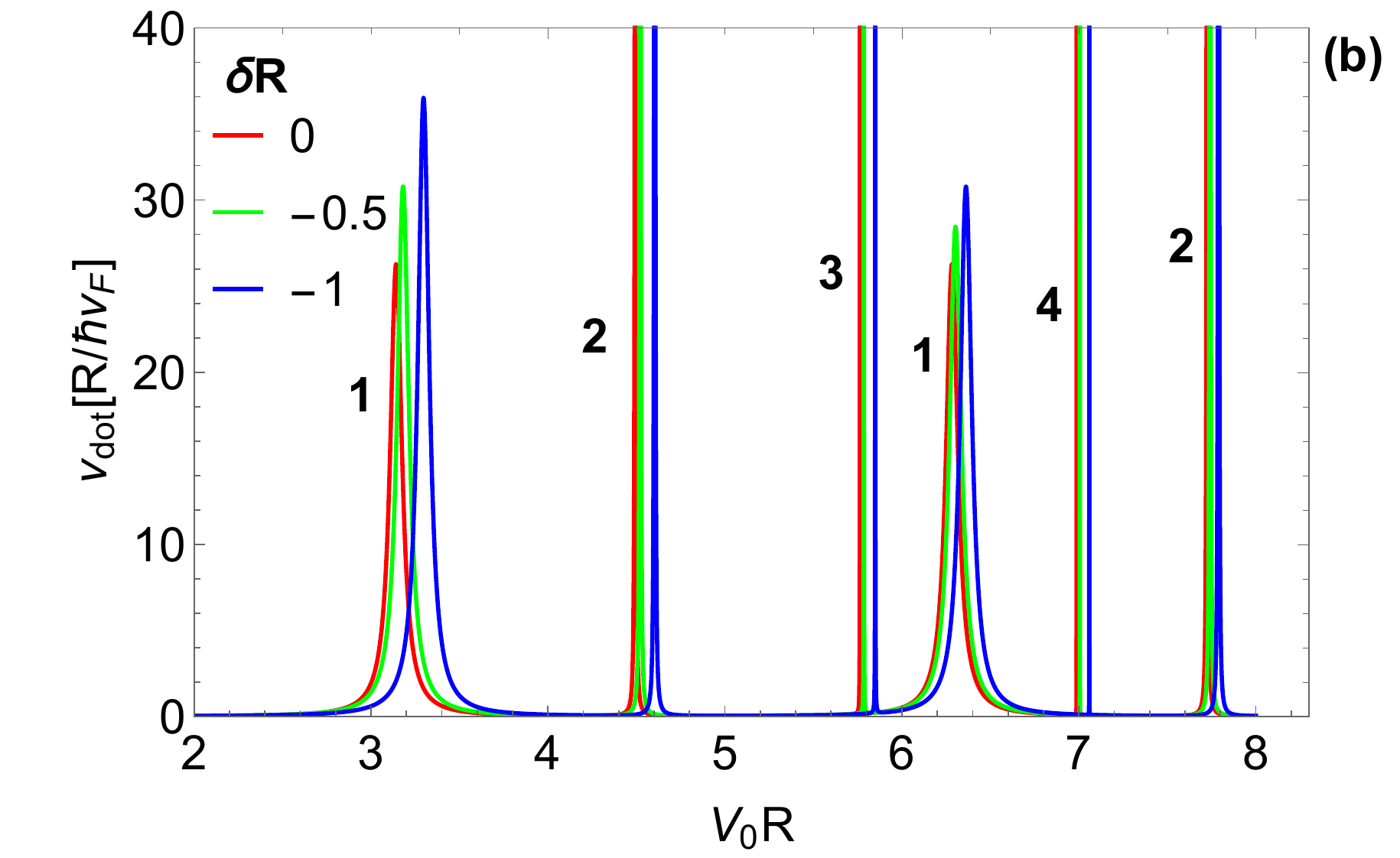}\includegraphics[width=0.5\linewidth]{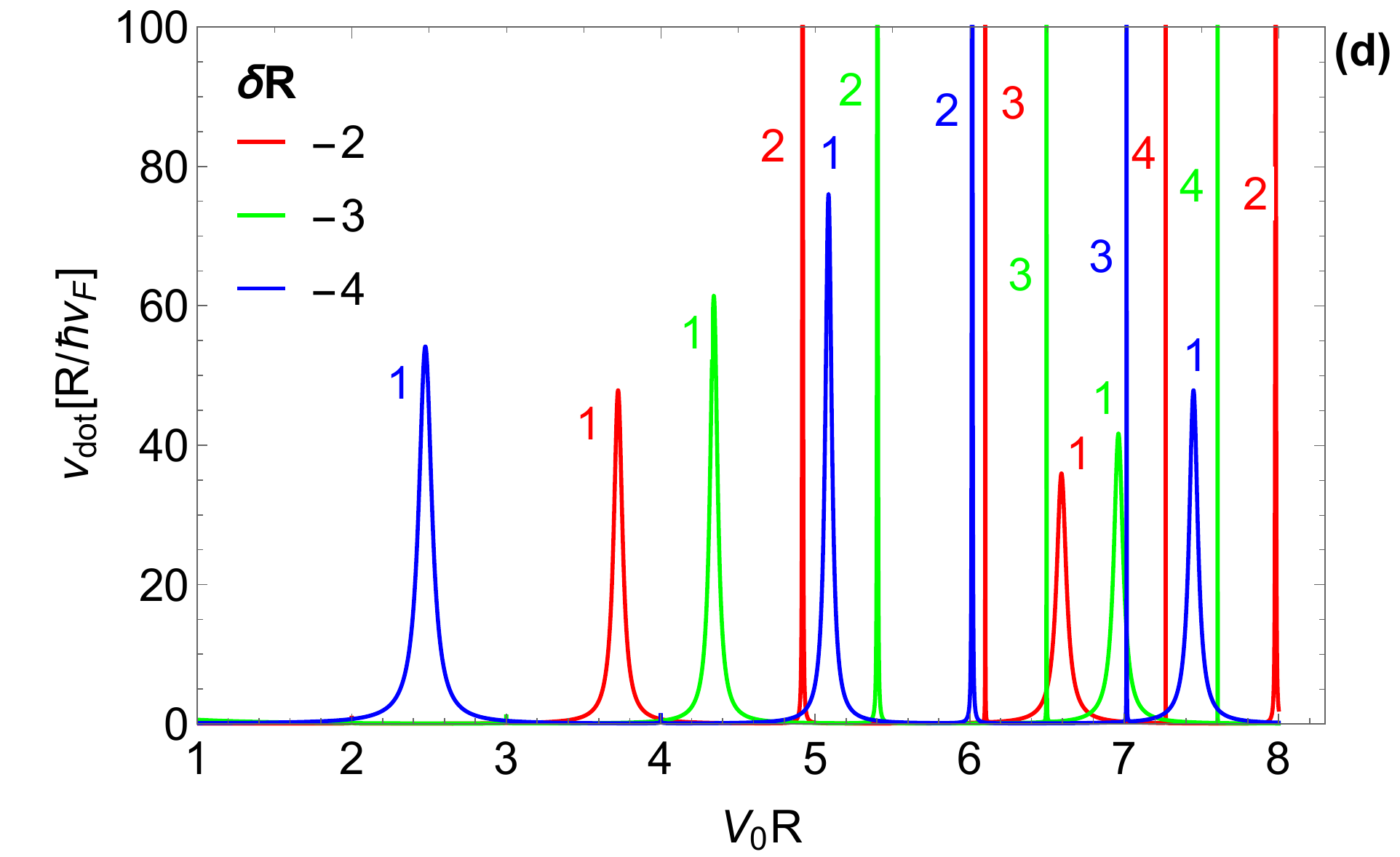}
\caption{\sf(color online) The DOS as function of the gate voltage $V_0 R$ at incident energy  $\epsilon=0$ and magnetic flux $\phi=1/2$ for $R/L=0.2$ and  different values of the energy gap $\delta R$. Here the resonances are labeled according to their  angular momentum $\mu=m+1/2$, with $\mu=\pm 1, \cdots,\pm 4$. (a): $\delta R =0, 0.5, 1$. (b): $\delta R=0, -0.5, -1$. (c): $\delta R =2, 3, 4$. (d): $\delta R=-2, -3, -4$.}
\label{f5}
\end{figure}

In Figure \ref{f5}, we show the DOS as a function of the gate voltage $V_0 R$ for $R/L=0.2$ and different values of $\delta R$ where the  resonances are labeled according to their angular momentum $\mu=m+1/2$. We observe that the amplitude of DOS decreases
as we increase $\delta R$ with a shift to the right for $\delta R$ positive (Figure \ref{f5}(a,c)) and a shift to left for $\delta R$ negative (Figure \ref{f5}(b,d)). The presence of the magnetic flux causes the elimination of the resonance corresponding to $m=-1/2$, that is, no normalizable bound state exists for this value of $m$. \\

\begin{figure}[h]\centering
	\includegraphics[width=0.5\linewidth]{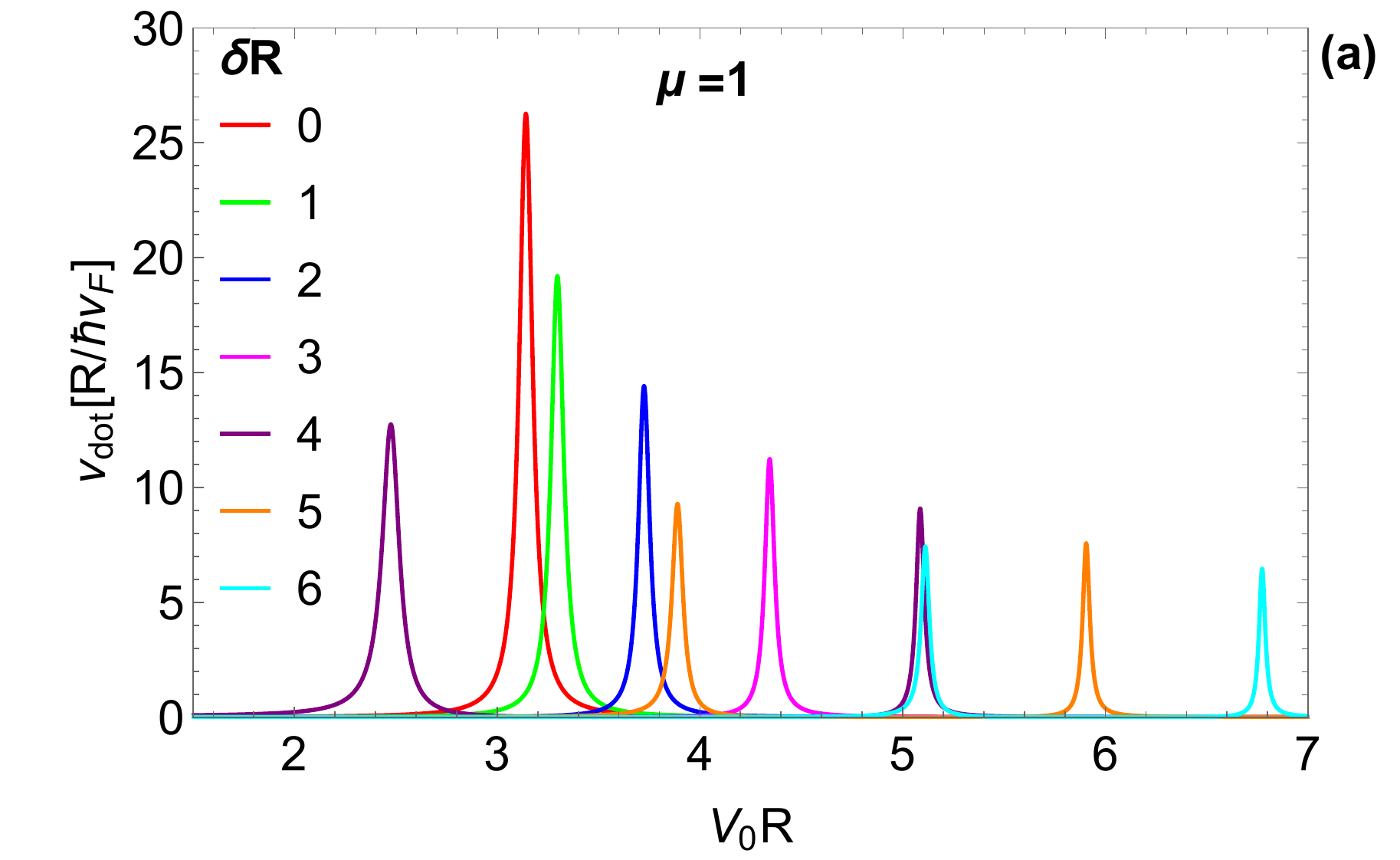}\includegraphics[width=0.5\linewidth]{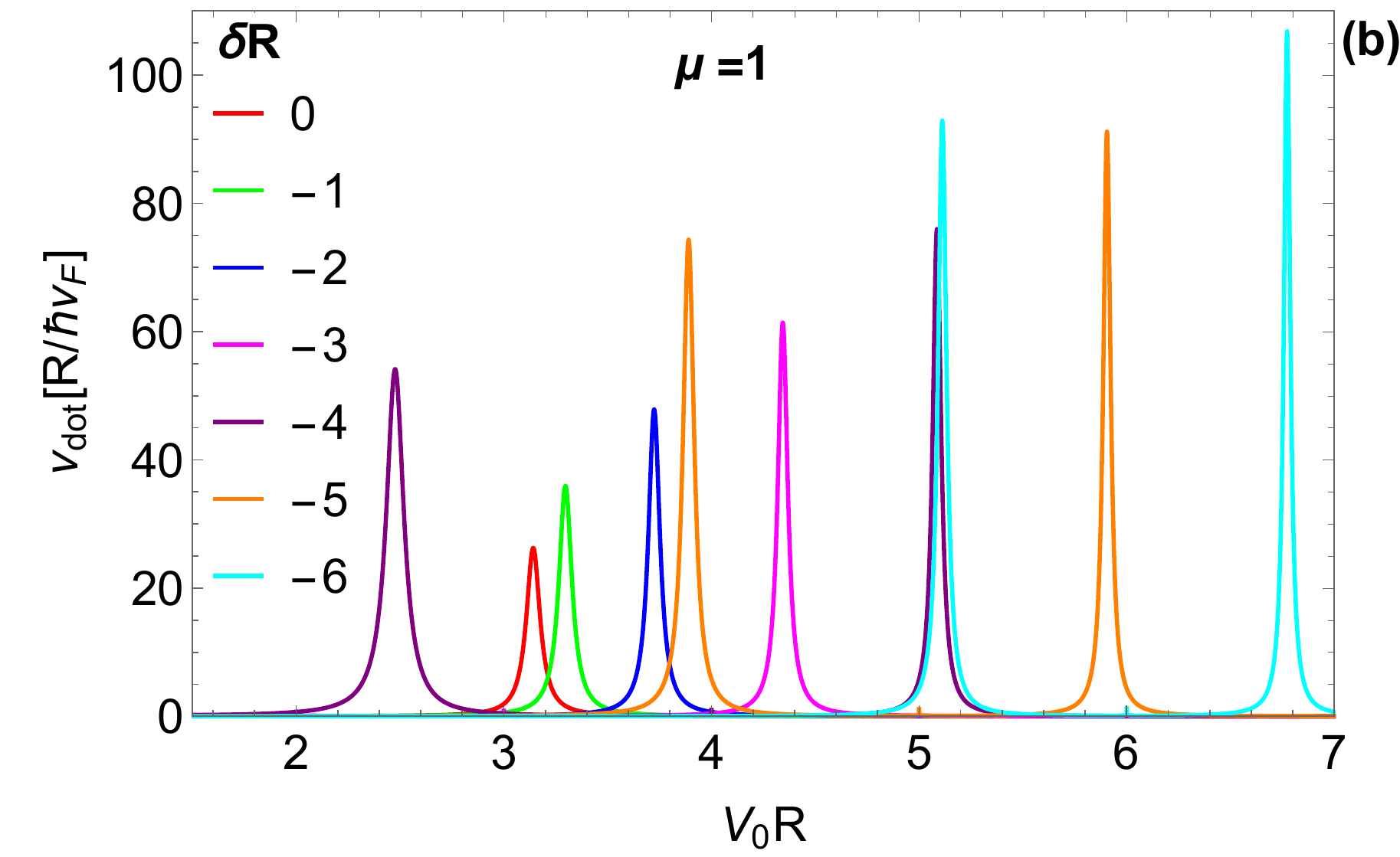}\\
	\includegraphics[width=0.5\linewidth]{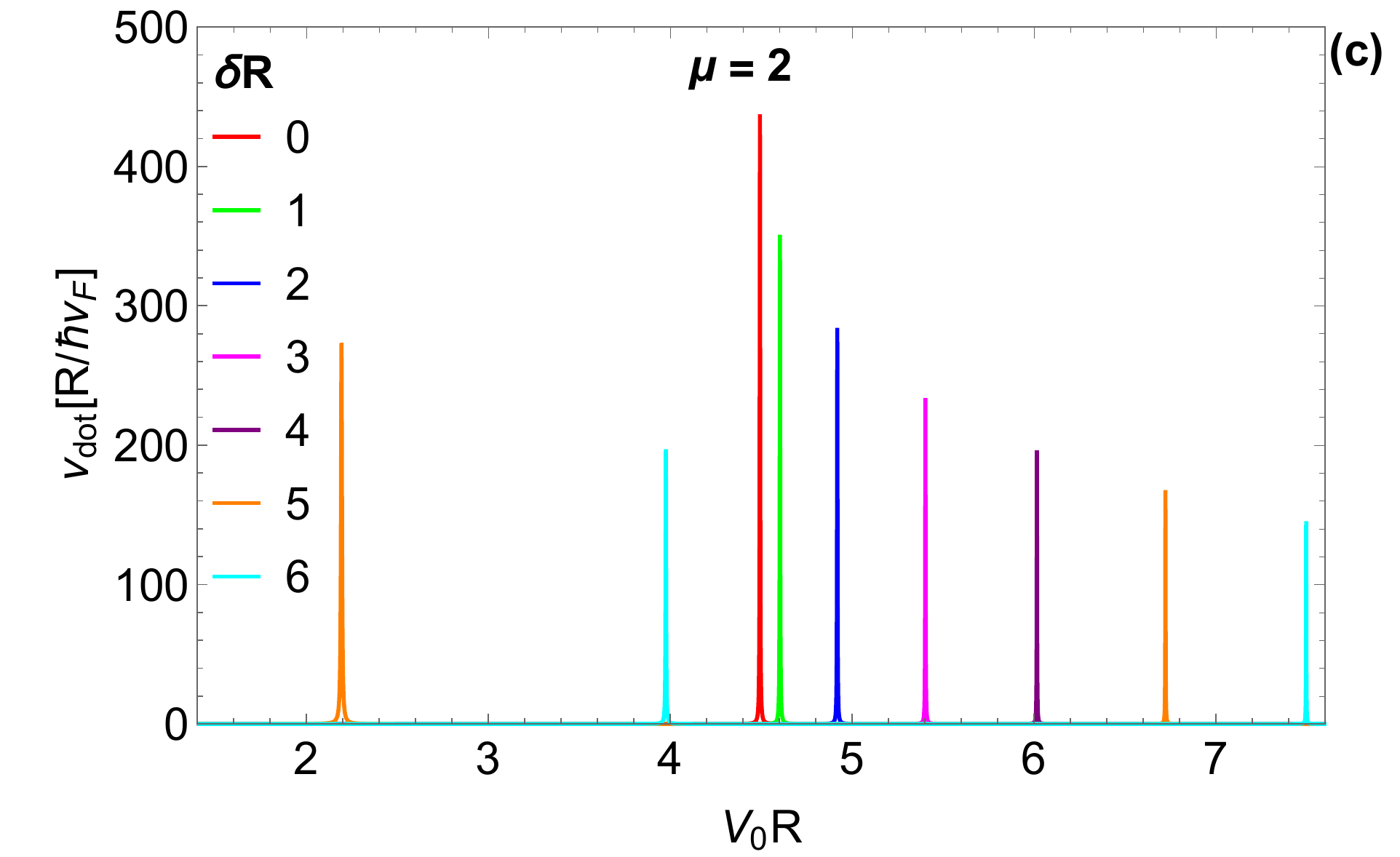}\includegraphics[width=0.5\linewidth]{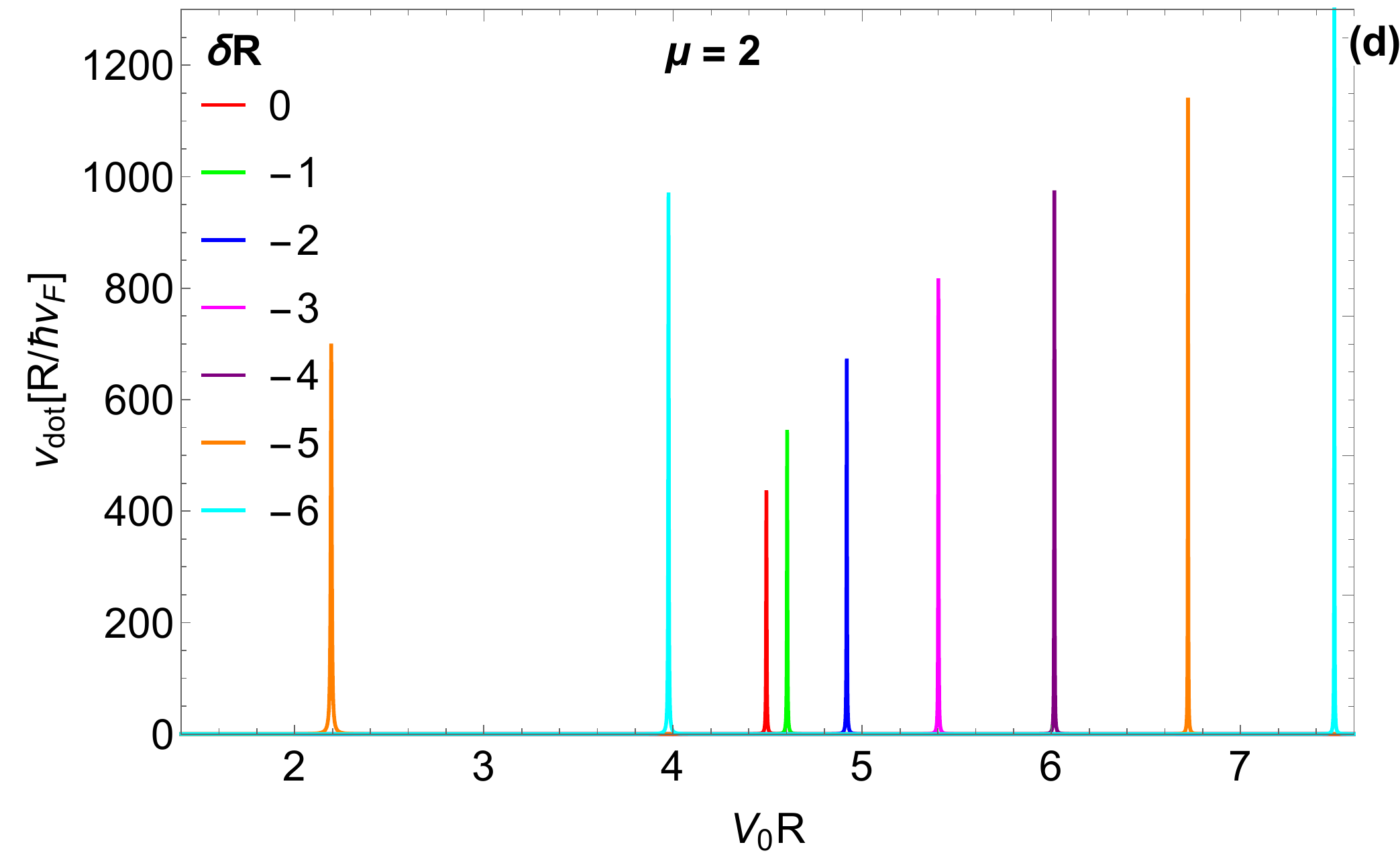}
	\caption{\sf (color online) The DOS as  function of the gate voltage $V_0 R$ at incident energy  $\epsilon=0$
		and magnetic flux  $\phi=1/2$ for $R/L=0.2$ and different values of the energy gap $\delta R$. (a): $\delta R =0, 1, 2, 3, 4, 5, 6$ and (b): $\delta R =0, -1, -2, -3, -4, -5, -6$ for fist resonance $\mu=1$. (c): $\delta R=0, 1, 2, 3, 4, 5, 6$ and (d): $\delta R =0, -1, -2, -3, -4, -5, -6$ for second resonance $\mu=2$.}\label{f6}
\end{figure}

In Figure \ref{f6}, we show the DOS as  function of the gate voltage $V_0 R$ in the presence of the magnetic flux $\phi=1/2$ at incident energy $\epsilon=0$ for $R/L=0.2$ and  different value of the energy gap $\delta R$. Figure \ref{f6}(a) corresponds to $\delta R=0, 1, 2, 3, 4, 5, 6$ with $\mu=1$ (first resonance), which shows that  
the DOS exhibits oscillations whose amplitudes decrease by increasing the enrgy gap $\delta R$. In addition, 
there is a doubling of the peaks as compared to the situation with $\delta R=4$. In Figure \ref{f6}(b) we choose  the values $\delta R=0, -1, -2, -3, -4, -5, -6$ with $\mu=1$ (first resonance), one sees that there is  the same behavior as that of Figure \ref{f6}(a) except that when the absolute value of $\delta R$ increases the amplitude of DOS increases. For the values $\delta R=0, 1, 2, 3, 4, 5, 6$ and $\mu=2$ (second resonance), Figure \ref{f6}(c) shows the appearance of peaks for each value of $\delta R$. The height of the peaks decreases when $\delta R$ increases, we also notice a doubling of resonances as compared to the situation with $\delta R=5$. Now for  $\delta R=0, -1, -2, -3, -4, -5, -6$, with $\mu=2$ (second resonance), Figure \ref{f6}(d) presents  the same behavior as that of Figure \ref{f6}(c) except that when the absolute value of $\delta R$ increases the oscillation amplitudes increase. Note that if we compare Figure \ref{f3} (absence of magnetic flux) and Figure \ref{f6} (presence of magnetic flux), we conclude that the energy gap in the presence of magnetic flux increases the heights and decreases the widths of the oscillation resonances.

\begin{figure}[h]\centering
	\includegraphics[width=0.33\linewidth]{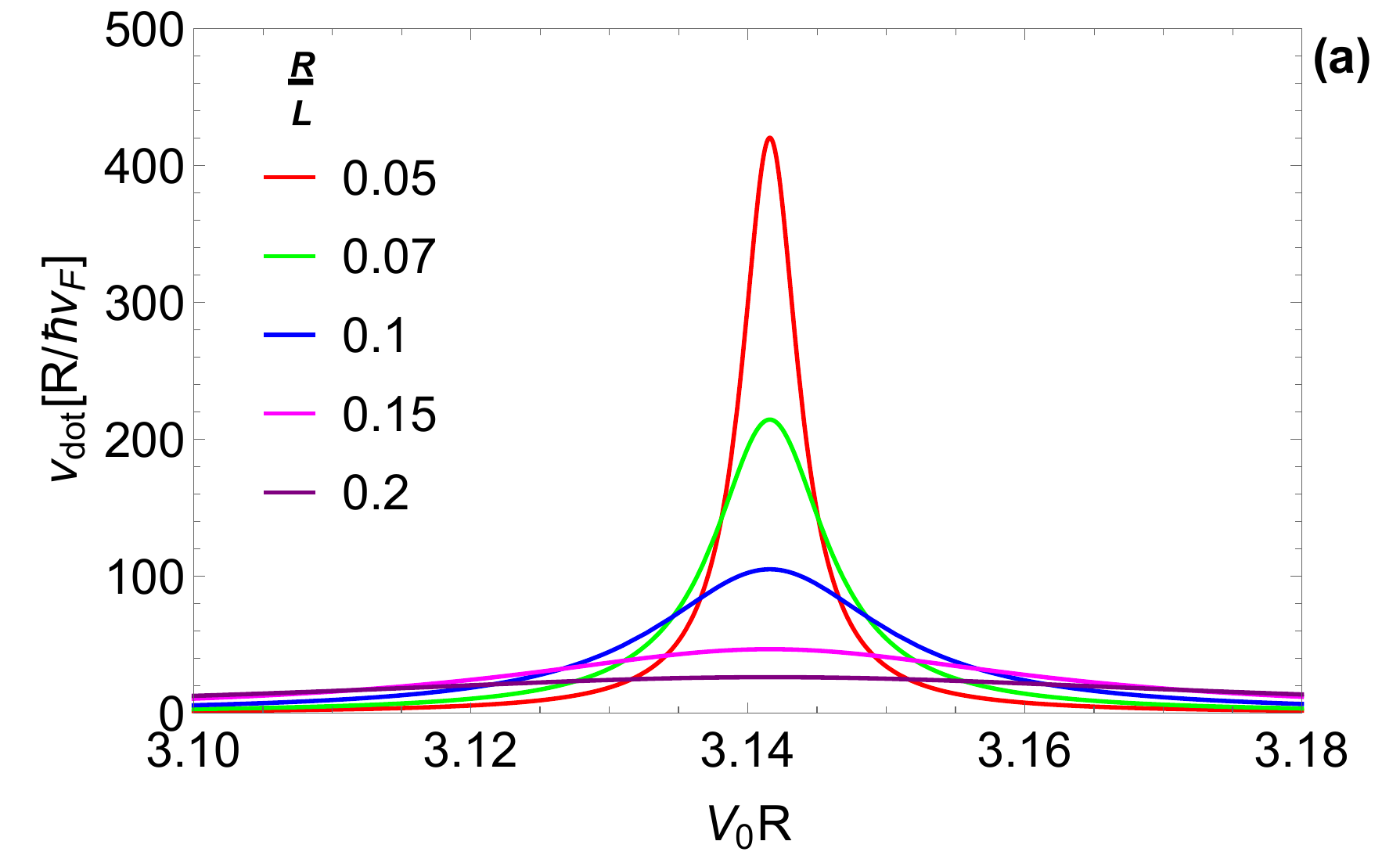}\includegraphics[width=0.33\linewidth]{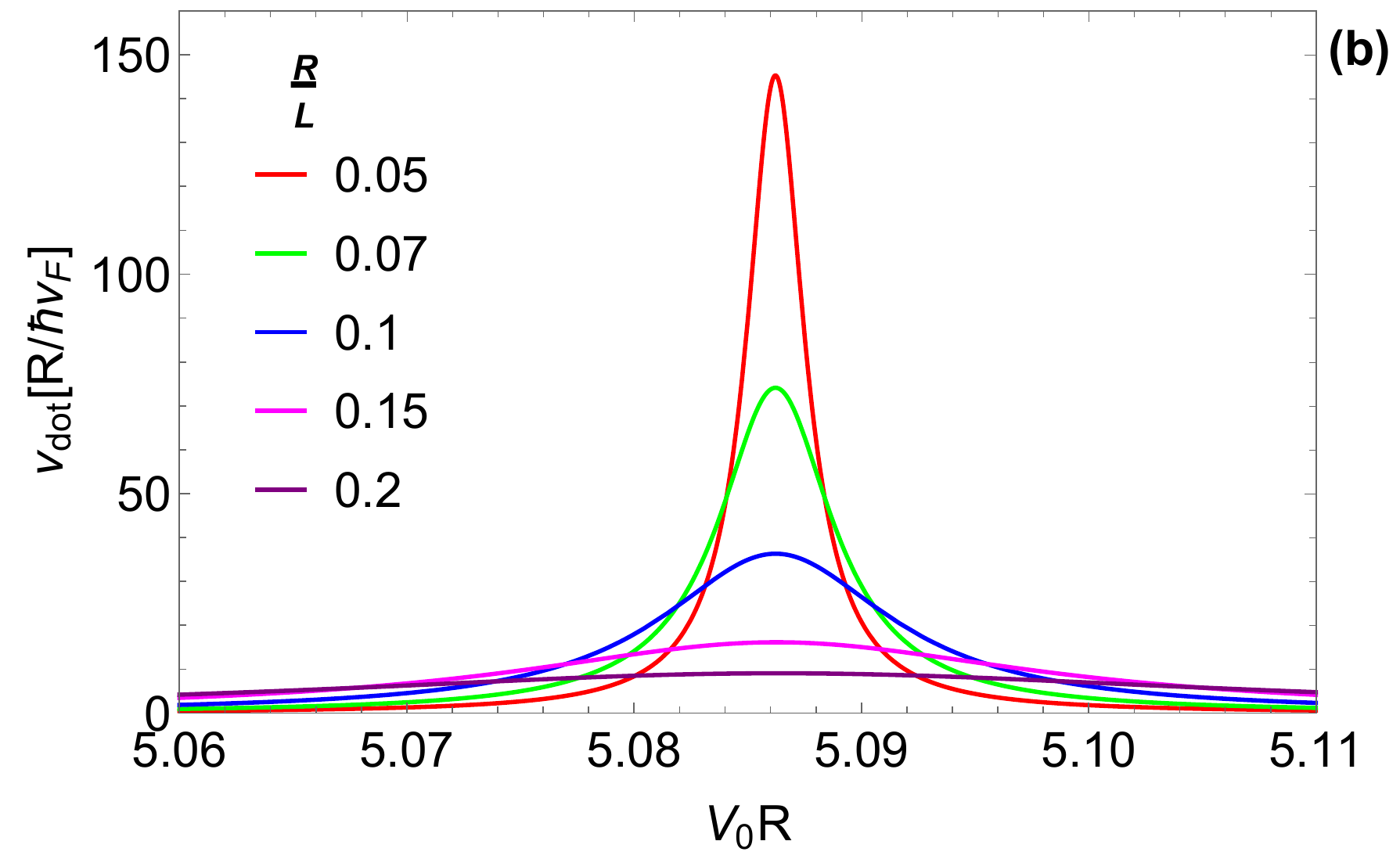}\includegraphics[width=0.33\linewidth]{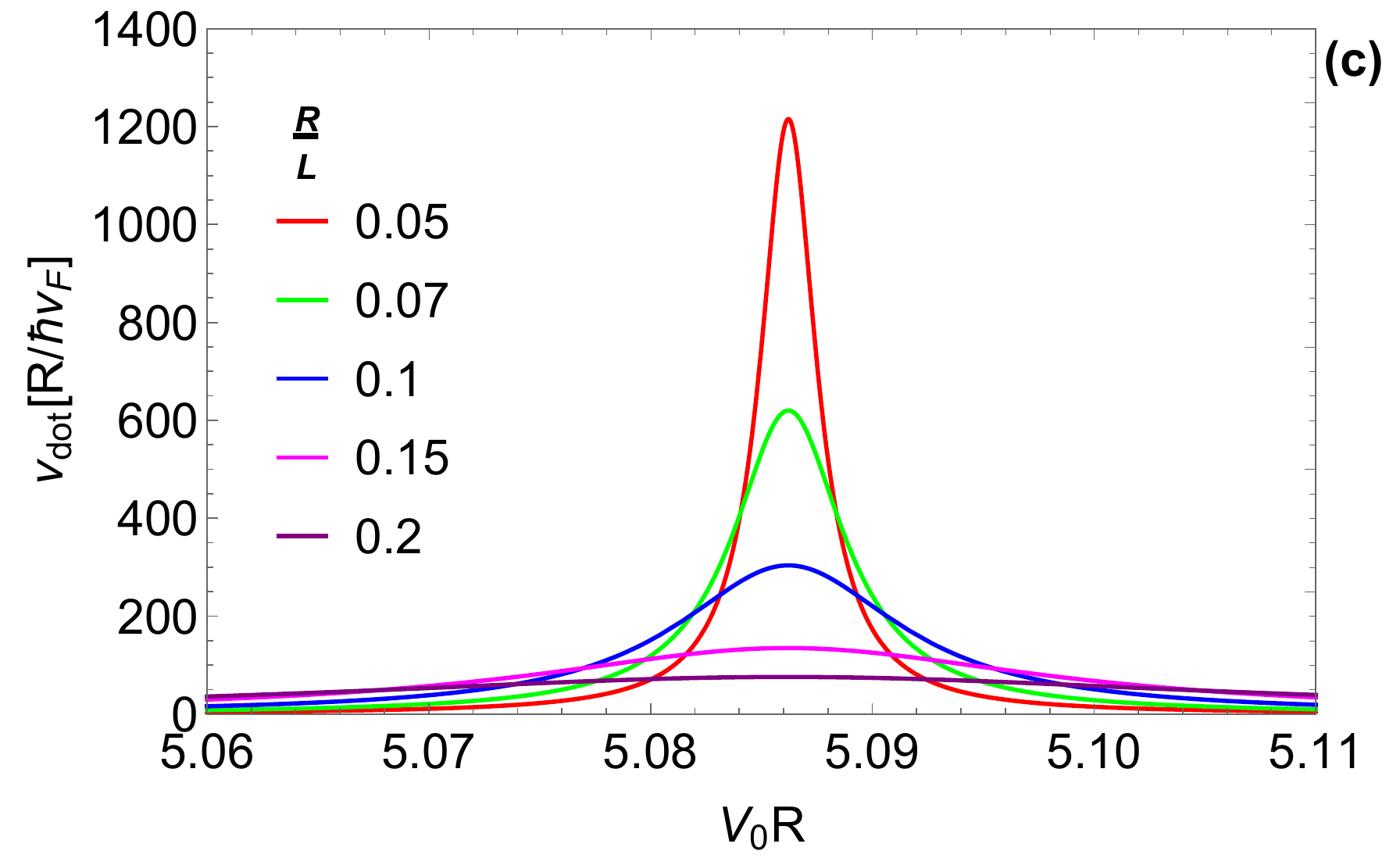}
	\caption{\sf(color online) The DOS  as function of the gate voltage $V_0 R$ at incident energy  $\epsilon=0$ and first resonance $\mu=1$ for different values of the energy gap $\delta R$ and  ratio $R/L$. (a):  $\delta R=0$, (b):  $\delta R=4$ and (c):  $\delta R=-4$ }\label{f7}
\end{figure}

In Figure \ref{f7}, we show how the first resonance $\mu=1$ behaves when we modify the energy gap $\delta R$ at $\epsilon=0$ and $\phi=1/2$. Indeed, the plot shows that the resonance width increases, while the height decreases when the contact size $L$ increases
for a fixed value of $R$. From Figure \ref{f7}(b,c), we notice that the introduction of $\delta R$ in the presence of  magnetic flux allows to amplify the resonance height by a factor of 10 as compared to
Figures \ref{f4}(b,c) for $m=1/2$ (zero magnetic flux).\\

In comparison to the DOS analysis reported in \cite{Martin14}, we have 
some comments in order. Indeed, 
we observe that considering an  energy gap $\delta$ in graphene quantum dot of radius $R$ with magnetic flux $\phi=1/2$ changes the resonance properties of the DOS. More precisely, we notice that  the amplitudes and  widths of resonances  decrease for the case $\delta>0$, 
but   they increase otherwise as well as  the  positions of resonances undergo changes. 
In addition, we observe that there are  appearance of the resonances and peaks when 
$\delta$ is greater than critical value, which can be fixed according to each considered configuration of the physical parameters.
In summary, the energy gap $\delta$ amplifies the DOS in the presence of magnetic flux and therefore we conclude that  it can be used as a tunable parameter to control the properties of our system. Of course the DOS results obtained in \cite{Martin14} can be recovered by  switching off $\delta$.

\section{Conclusion}

We have studied the confinement of charge carriers in a quantum dot of  graphene
surrounded by a sheet of undoped graphene and connected to a metallic contacts in the presence of an energy gap and magnetic flux.
We have solved the two-band Dirac Hamiltonian in the vicinity of the $K$ and $K'$ valleys and obtained analytically the solutions of energy spectrum for three regions composing our system.
Using the asymptotic behavior of the 
Hankel functions for large arguments, we have derived an approximate formula for the the density of states (DOS) as a function of magnetic flux, energy gap and the applied electrostatic potential. We have
found the resonance conditions at zero energy under suitable boundary conditions.

We have shown that the DOS exhibits an oscillatory behavior which reflects the
appearance of resonances. The amplitude of DOS oscillation resonances was found to decrease and shift to the right when $\delta$ increases for $\delta>0$. On the other hand, when $\delta$ is negative the resonance peaks shift to the left. It was also observed that
for higher values of the angular momentum $m$, the resonances disappear and peaks take places either in presence or absence of magnetic flux.
We have shown that the presence of magnetic flux eliminates the resonance which correspond to $m=-\frac{1}{2}$ while the resonances corresponding to  $m\neq -\frac{1}{2}$ becomes sharper with an amplification of its amplitude.

\section*{Acknowledgments}
The generous  support provided by the Saudi Center for Theoretical Physics (SCTP) is highly appreciated by all authors.
 AJ and HB also acknowledge the support of King Fahd University of Petroleum and Minerals under research group project RG181001.

\end{document}